\documentclass[fleqn,usenatbib]{mnras}

\usepackage{newtxtext,newtxmath}

\usepackage[T1]{fontenc}

\DeclareRobustCommand{\VAN}[3]{#2}
\let\VANthebibliography\thebibliography
\def\thebibliography{\DeclareRobustCommand{\VAN}[3]{##3}\VANthebibliography}

\usepackage{subfigure}

\usepackage{graphicx}	
\usepackage{amsmath}
\usepackage{amssymb}	

\title[Evolution of magnetic fields in stars] { Evolution of random initial magnetic fields in stably stratified and barotropic stars}

\author[Becerra et al.]{
Laura Becerra,$^{1}$\thanks{E-mail: lbecerra@astro.puc.cl}
Andreas Reisenegger,$^{2}$
Juan Alejandro Valdivia,$^{3,4}$
Mikhail~E.~Gusakov$^{5}$
\\
$^{1}$Instituto de Astrof\'isica, Facultad de F{\'i}sica, Pontificia Universidad Cat\'olica de Chile, Av. Vicu\~na Mackenna 4860, Macul, Santiago, Chile\\
    $^{2}$Departamento de F\'{\i}sica, Facultad de Ciencias B\'asicas, Universidad Metropolitana de Ciencias de la Educaci\'on, Av. Jos\'e Pedro Alessandri 774,\\ \~Nu\~noa, Santiago, Chile\\
$^{3}$Departamento de F\'{\i}sica, Facultad de Ciencias,
Universidad de Chile, Las Palmeras 3425, \~Nu\~noa, Santiago, Chile\\
$^{4}$Centro para el Desarrollo de la Nanociencia y
Nanotecnolog\'\i a, CEDENNA, Santiago, Chile\\
$^{5}$Ioffe Institute, 26 Politekhnicheskaya Street, St. Petersburg 194021, Russia
}

\date{Accepted XXX. Received YYY; in original form ZZZ}

\pubyear{2021}

\begin{document}
\label{firstpage}
\pagerange{\pageref{firstpage}--\pageref{lastpage}}
\maketitle

\begin{abstract}

Long-lived magnetic fields are known to exist in upper main-sequence stars, white dwarfs, and neutron stars. In order to explore possible equilibrium configurations of the magnetic field inside these stars, we have performed 3D-magnetohydrodynamic simulations of the evolution of initially random magnetic fields in stably stratified and barotropic stars with an ideal-gas equation of state using the {\sc Pencil Code}, a high-order finite-difference code for compressible hydrodynamic flows in the presence of  magnetic fields. In barotropic (isentropic) stars, we confirm previous results in the sense that all initial magnetic fields we tried decay away, unable to reach a stable equilibrium. In the case of stably stratified stars (with radially increasing specific entropy), initially random magnetic fields appear to always evolve to a stable equilibrium. However, the nature of this equilibrium depends on the dissipation mechanisms considered. If magnetic diffusivity (or hyperdiffusivity) is included, the final state is more axially symmetric and dominated by large wavelengths than the initial state, whereas this is not the case if only viscosity (or hyperviscosity) is present. In real stars, the main mechanism allowing them to relax to an equilibrium is likely to be phase mixing, which we argue is more closely mimicked by viscosity. Therefore, we conclude that, depending on its formation mechanism, the equilibrium magnetic field in these stars could in principle be very asymmetric.
\end{abstract}




\section{Introduction}

Long-lived magnetic fields have been observed in a wide variety of stars, from the pre-main sequence to white dwarfs and neutron stars. About 10\% of the population of intermediate-mass stars (from $1.5~M_\odot$ to $8~M_\odot$) host steady and globally organized magnetic fields \citep{2009ARA&A..47..333D}. They have been observed, specially, in the subgroup of chemically peculiar main-sequence stars (Ap/Bp stars) with surface magnetic field strengths ranging from 300~G to 30~kG \citep{2007A&A...475.1053A,2019MNRAS.483.3127S}.  In general, these are modeled as dipole fields with their magnetic axis not aligned with their rotational axis \citep{2000A&A...359..213L}, but magnetic Doppler imaging studies based on full Stokes vector spectropolarimetric observations have revealed more complex field geometries \citep{2000MNRAS.313..823W,2020pase.conf...89K}. 

Strong magnetic fields similar to those of Ap/Bp stars also appear in $\sim 7$\% of the more massive OB stars \citep{2017MNRAS.465.2432G}. Moreover, magnitude-limited surveys reveal that around $\sim 3$ \% of the isolated white dwarfs have a large-scale magnetic field with strength between $\sim 10^6$~G and $\sim 10^9$~G, but the incidence of weakly magnetic white dwarfs with fields $<10^6$~G is still uncertain \citep{2020AdSpR..66.1025F}. In neutron stars, a wide range of magnetic fields can be found, from $\sim 10^8$~G in millisecond pulsars and low-mass X-ray binaries to $\sim 10^{15}$~G in magnetars \citep{2015SSRv..191..315M, 2017ARA&A..55..261K}.

The stars in question either have stably stratified (``radiative'') envelopes (intermediate and massive main-sequence stars) or are stably stratified throughout their interior (white dwarfs and neutron stars). This and the observed stability of these magnetic fields over long timescales favor a fossil field origin, namely the magnetic field is not continuously renewed, but instead it was formed in a previous evolutionary stage or event in the star's past \citep{1945MNRAS.105..166C,2001ASPC..248..305M}.  However, it remains unclear how and at which evolutionary stage the seed magnetic field was formed. The detection of strong and organized magnetic fields in pre-main-sequence Herbig Ae/Be stars \citep{2013MNRAS.429.1027A} and the observation of magnetic fields in post-main-sequence red giants stars \citep{2008A&A...491..499A,2017MNRAS.471.1926N} may provide important constraints to the origin and evolutionary path of magnetic stars.

Since observations only give (partial) information about the surface magnetic fields, finding equilibrium magnetic field configurations inside the star and proving their stability over long timescales, much longer than the dynamical timescale, is important to the fossil field theory, but also to determine the influence of the magnetic field on the stellar structure and evolution. This problem has been a matter of research since \cite{1953ApJ...118..116C}, and important advances have been made since then. For example, based on an energy principle method \citep{1958RSPSA.244...17B}, it was formally shown that both purely toroidal and purely poloidal magnetic fields become unstable to adiabatic perturbations somewhere in the star \citep{1973MNRAS.161..365T, 1973MNRAS.163...77M,1973MNRAS.162..339W}. The instability growth rate  is of the order of the Alfv\'en frequency, i.~e., $\sim [10$~yr]$^{-1}$ in a main-sequence star with a field of $1$~kG, $\sim$ days$^{-1}$ in a white dwarf with $10$~MG, and $\sim [100$~s]$^{-1}$  for neutron stars with $10^{12}$~G.

\citet{2006A&A...450.1077B} \cite[see also][]{2004Natur...431..819B}  evolved disordered initial magnetic fields numerically in time and found that, over a few Alfv\'en timescales, they relaxed to stable, roughly axisymmetric equilibria consisting of both toroidal and poloidal components of comparable strength in a twisted-torus shape, compatible qualitatively with the suggestion of \cite{1956ApJ...123..498P}.  \citet{2008MNRAS.386.1947B} found that a random initial field could also relax into a non-axisymmetric equilibrium, depending on the initial conditions: a centrally concentrated field evolves into a roughly axisymmetric equilibrium, and a more spread-out magnetic field evolves into a more complex geometry.

It has been argued that an essential ingredient for the stability of the magnetic field is the stable stratification of the matter inside the star \citep{2009MNRAS.397..763B,2013MNRAS.433.2445A}. In fact, on short timescales, the radiative envelopes of massive stars and the interiors of white dwarfs are stabilized by entropy gradients, while in neutron star cores this role is played by a varying chemical composition \citep{2009A&A...499..557R}. \cite{2012MNRAS.424..482L} and \cite{2015MNRAS.447.1213M} tested the stability of a wide variety of mixed toroidal-poloidal axisymmetric fields in barotropic (i.~e., non-stably stratified) stars and did not find any stable configurations. \cite{2015MNRAS.447.1213M}  \citep[see also][]{2012MNRAS.422..619B} additionally evolved initially disordered magnetic fields in barotropic stars, finding that the magnetic field never evolved into a stable configuration.

In the present paper, we focus on the early evolution of the magnetic field by simulating the evolution of an initially random configuration; as in \cite{2004Natur...431..819B,2006A&A...450.1077B}, and \cite{2015MNRAS.447.1213M}; but going beyond these works in various aspects. 

First, all those previous papers used the same numerical code, namely the {\sc Stagger code} of \citet{2005ApJ...618.1020G}. Here, we use the publicly available {\sc Pencil Code} \citep{2021JOSS....6.2807P}, a high-order finite-difference code for compressible flows that has previously been used in turbulence simulations, accretion disk outflows, and dynamo experiments, in order to independently check the main results of the previous simulations. We confirm that stable stratification plays a crucial role, in the sense that initial random fields evolve into stable equilibria if the stellar matter is stably stratified, whereas they decay away in the case of barotropic stars.

Second, we explore the effect of different dissipation mechanisms. At the formation of a star or after a violent episode in its evolution, such as a binary interaction, the star will likely be in a non-equilibrium state with a disordered magnetic field. Force imbalances will either cause instabilities leading to the loss of the magnetic field from the star or the generation of sound and Alfv{\'e}n waves traveling through the star, which are likely damped by phase mixing \citep{1999A&A...349..189S}, allowing the star to settle into a stable hydromagnetic equilibrium state, in which the Lorentz force is balanced by pressure and gravity forces.

Phase mixing means that Alfv{\'e}n waves traveling on neighboring field lines will quickly get out of phase with each other, creating small-scale gradients that are damped by viscosity or magnetic diffusivity. Numerical simulations cannot reach the fine resolution to model this process in any realistic way, which furthermore would require a huge dynamical range in timescales (fast wave propagation and very slow damping). Thus, they must rely on other, unrealistically strong dissipation mechanisms in order to qualitatively mimic its effects. 

The previous simulations mentioned above include high-order ``hyper-diffusion'' terms in the time-evolution equations for velocity, magnetic field, and specific entropy, which have the advantage of preventing instabilities by smoothing the smallest scales, but presumably with little effect on the larger structures. Here, we run simulations with different kinds of dissipation (i.~e., ordinary viscosity, hyper-viscosity, magnetic diffusion, and magnetic hyper-diffusion) to elucidate their specific effects on the magnetic field evolution, finding that magnetic diffusion processes make the field become more ordered and axisymmetric, while viscous processes do not. 
We argue that the latter are closer analogs to phase mixing, as they damp waves, but do not dissipate an equilibrium magnetic field configuration. Thus, we conclude that the stable equilibria set up in real stars may well be quite disordered and asymmetric, of course depending on its formation mechanism, which sets the realistic initial conditions.

We also explore the effects of different set-ups for the initial magnetic field (central concentration and power spectrum), finding that they do not lead to qualitatively different outcomes.

In Section~\ref{sec:Eqs_code}, we present the MHD equations and the numerical set-up for the simulations, while in Section~\ref{sec:definitions}, we define the quantities used to analyze their outcomes. In Section~\ref{sec:PhyDampMch} we calculate the timescales of physical damping mechanisms that allow the magnetic field configuration to relax to an equilibrium inside stars and discuss how we model this process. In the subsections of Section~\ref{sec:sim_results}, we present the results of simulations of initial random magnetic fields evolving in model stars, exploring several variables while such as dissipation by (hyper-)viscosity (\S~\ref{sec:sim_visc}) and magnetic (hyper-)diffusion (\S~\ref{sec:sim_diffusivity}) inside the star, the magnetic diffusivity profile in the stellar atmosphere (\S~\ref{sec:sim_diffAtm}), the spatial structure of the initial field (\S~\ref{sim:initial_magconf}), and the hypothetical absence of stable stratification (\S~\ref{sec:sims_BarStar}). Our conclusions are presented in Section~\ref{sec:conclusion}.

\section{Equations and set-up}\label{sec:Eqs_code}

%
\begin{figure*}
    \centering
    \subfigure[Density]{\includegraphics[width=0.3\textwidth]{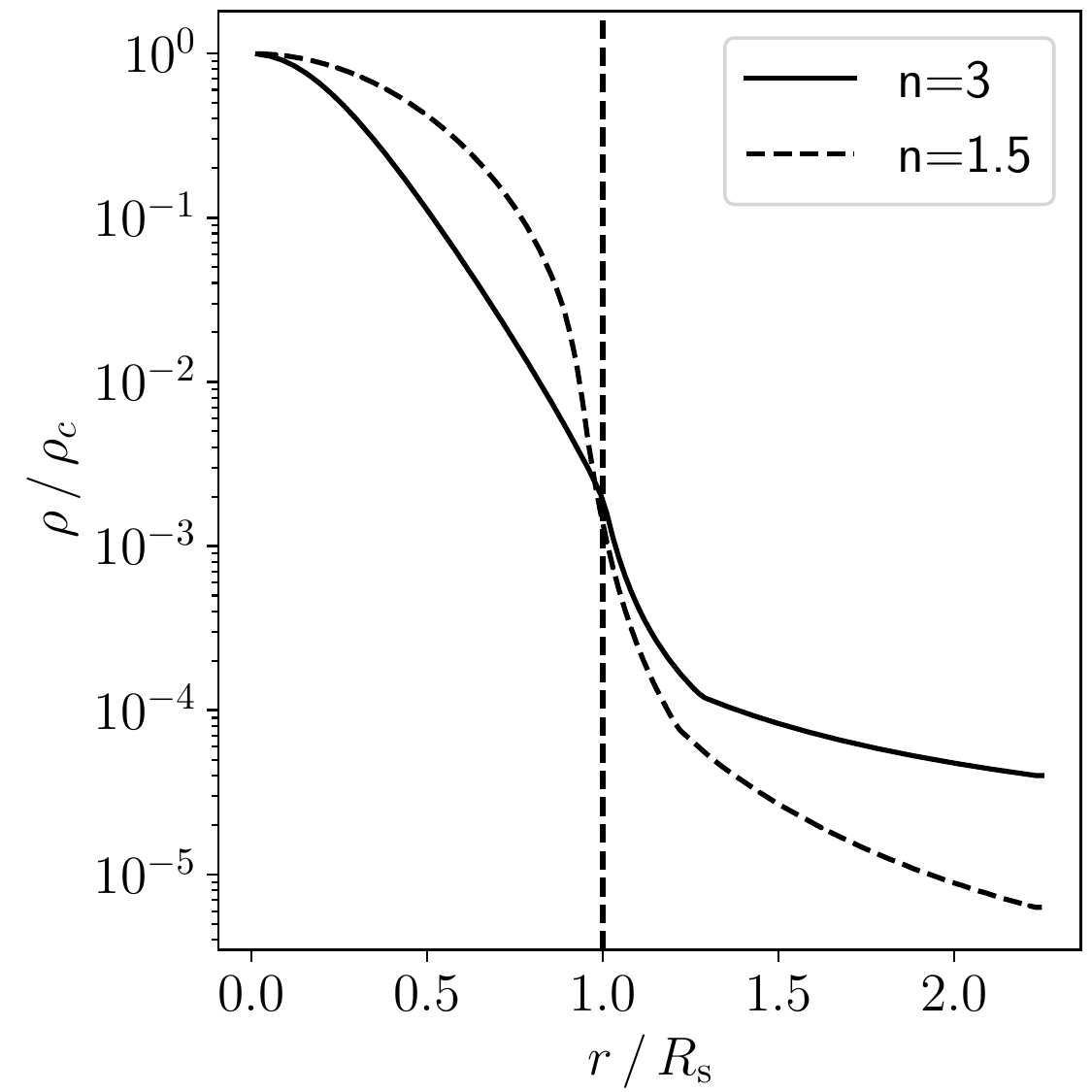}}
    \subfigure[Specific entropy]{\includegraphics[width=0.3\textwidth]{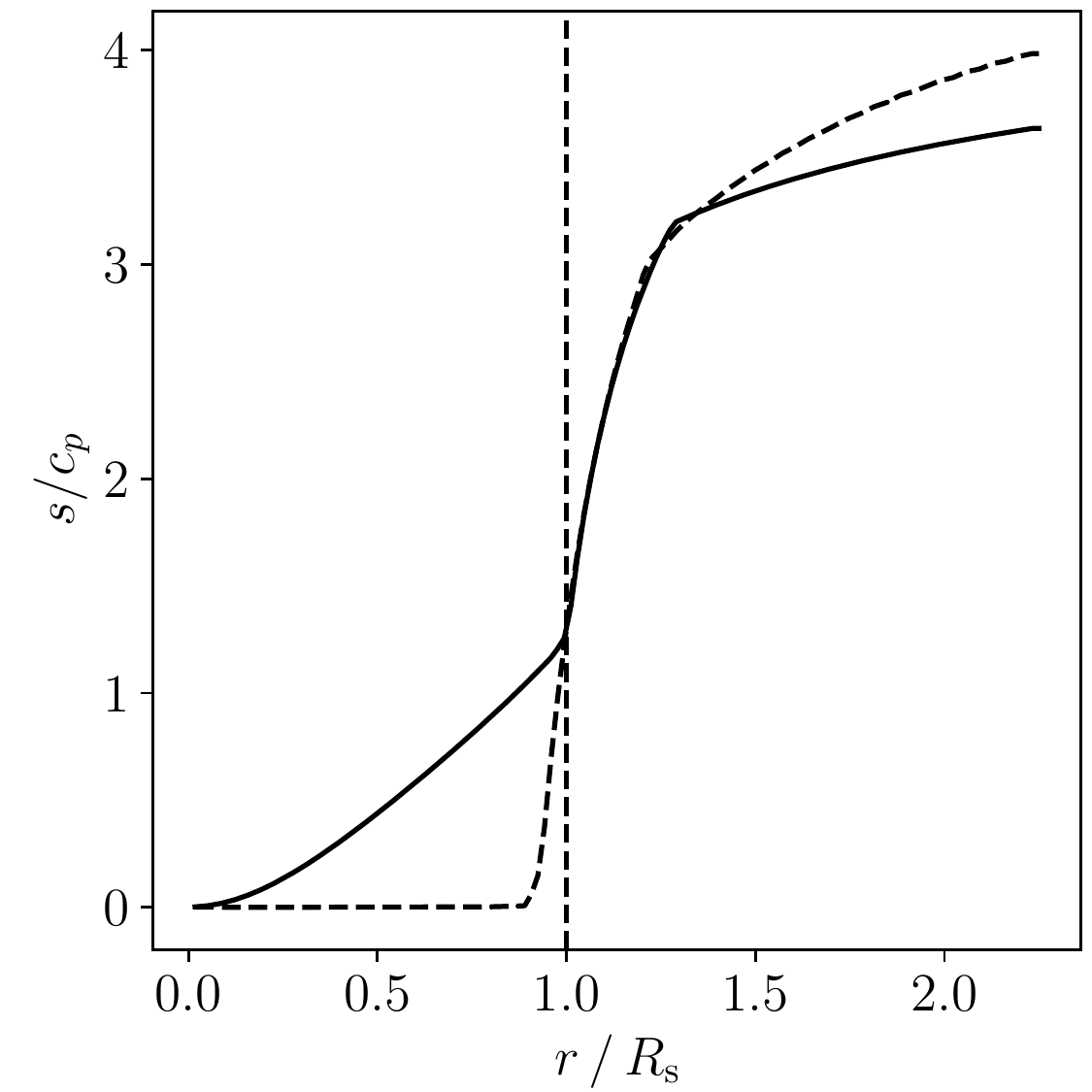}}
    \subfigure[Temperature]{\includegraphics[width=0.305\textwidth]{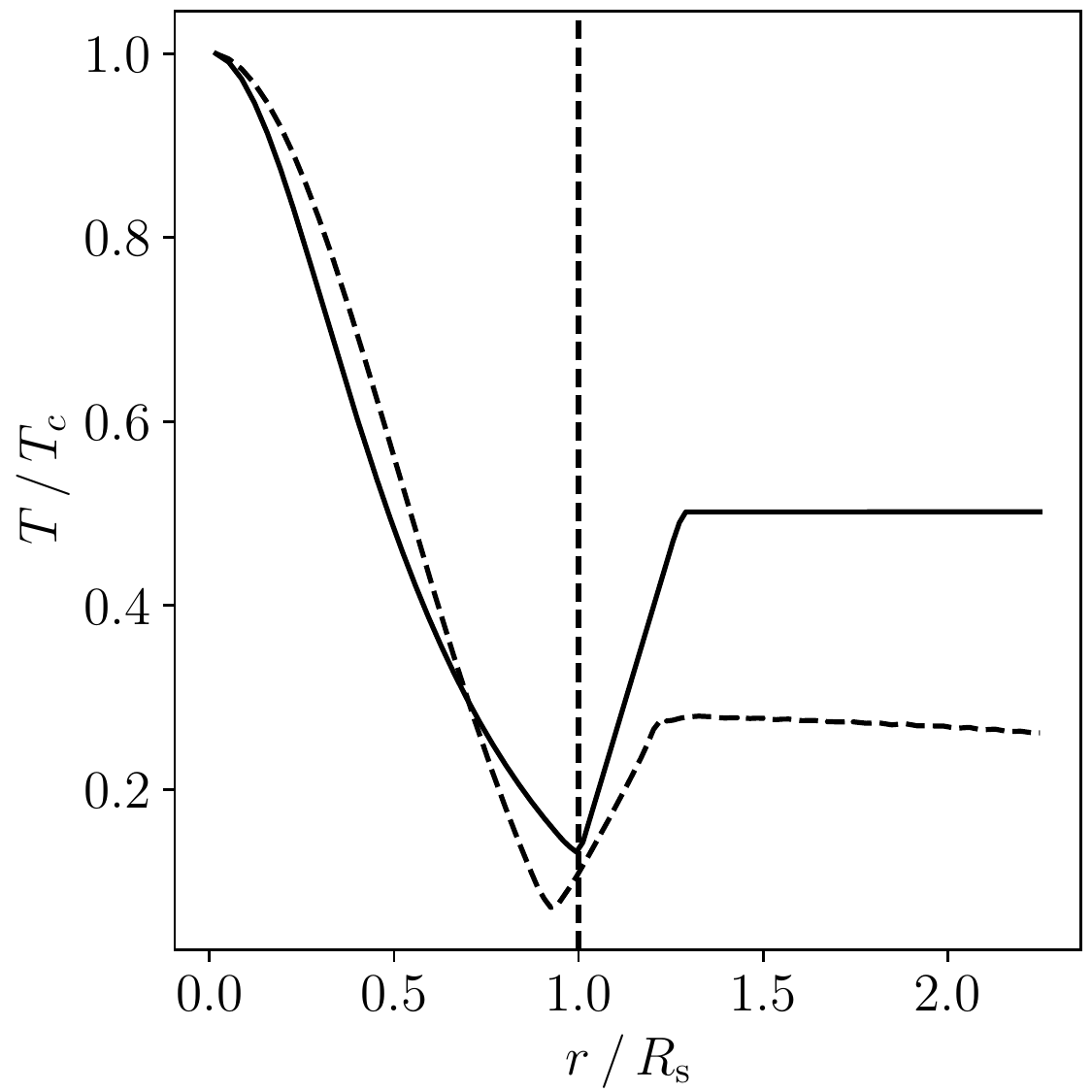}}
    \caption{Initial radial profiles for density, entropy, and temperature for a polytropic star with $n=3$ (solid line) and $n=1.5$ (dashed line).   
    }
    \label{fig:init_cond}
\end{figure*}

We solve the magneto-hydrodynamics (MHD) equations:
\begin{eqnarray}\label{eq:MHDEqs_a}
   \frac{\partial ({\rm ln}\,\rho)}{\partial t} &=& -\vec{\nabla}\cdot\vec{u} - \vec{u} \cdot \vec{\nabla}({\rm ln}\,\rho)  \\
   \label{eq:MHDEqs_b}
   \frac{\partial \vec{u}}{\partial t} &=&  -\vec{u}\cdot\vec{\nabla}\vec{u} - \frac{\vec{\nabla}p}{\rho}-\vec{\nabla}\Phi + \frac{\vec{j}\times\vec{B}}{\rho} + \vec{f}_{\rm visc}+  \vec{f}^{\rm hyper}_{\rm visc} \\
 \label{eq:MHDEqs_c}
   \frac{\partial\vec{A}}{\partial t} &=& \vec{u}\times\vec{B} -\eta \mu_0\vec{j} + \eta_3\nabla^6\vec{A}\\
   \label{eq:MHDEqs_d}
    \frac{\partial s}{\partial t} &=& -\vec{u}\cdot \vec{\nabla} s + \frac{\eta\mu_0|\vec{j}|^2}{\rho T} + \frac{2\nu\mathbf{S}^2}{T} + \mathcal{H}^{\rm hyper} \, ,
\end{eqnarray}
where $\rho$, $p$, $\vec{u}$,  and $s$ are the fluid mass density, pressure, velocity, and specific
entropy (per unit mass), respectively.  Here, $\Phi$ is the gravitational potential; $\vec{A}$ and $\vec{B}=\vec{\nabla}\times\vec{A}$ are the magnetic vector potential and magnetic field, respectively; $\eta$ is the magnetic diffusivity and $\vec{j}=\mu_0^{-1}\vec{\nabla}\times\vec{B}$ is the current density; $\mu_0$ is the magnetic vacuum permeability; and the viscous force is given by
\begin{equation}\label{eq:visc_force}
 \vec{f}_{\rm visc}=\rho^{-1}\vec{\nabla}\cdot\left(2\rho \nu{\mathbf{S}} \right) \, ,
\end{equation}
with  $\nu$, the kinematic viscosity, and $\mathbf{S}$, the rate-of-shear tensor, whose components are
\begin{equation}\label{eq:shear_tensor}
S_{ij}= \frac{1}{2}\left[\frac{\partial u_j}{\partial x_i}+ \frac{\partial u_i}{\partial x_j}-\frac{2}{3}\delta_{ij}(\vec{\nabla}\cdot\vec{u})\right]\, .
\end{equation}

The last terms in equations (\ref{eq:MHDEqs_b}) and (\ref{eq:MHDEqs_c}) correspond to an explicit sixth-order hyper-diffusion scheme added for numerical stability \citep[see][for examples of implementation]{2005ApJ...634.1353J, 2017AJ....154..146L}. The operator $\nabla^6\equiv\nabla^2(\nabla^2(\nabla^2))$ and the hyper-viscosity force is 
\begin{equation}\label{eq:F_visc}
    \vec{f}^{\rm hyper}_{\rm visc} = \rho^{-1}\vec{\nabla}\cdot\left(2\rho \nu_3 {\mathbf{S}}^{(3)}\right),\quad {\rm with} \quad   \mathbf{S}^{(3)}=(-\nabla^2)^2\,\mathbf{S}\, .
\end{equation}
The ``hyper-diffusion coefficients'' $\eta_3$ (``magnetic hyper-diffusivity'') and  $\nu_3$ (``hyper-viscosity''),  are constant. For self-consistency, the kinetic and magnetic energy dissipated by the hyper-viscosity force and magnetic hyper-diffusivity, respectively, have been added as heat sources in equation~(\ref{eq:MHDEqs_d}) through the term 
\begin{equation}\label{eq:heat_hyper}
    \mathcal{H}^{\rm hyper} = -\frac{\vec{u}\cdot \vec{f}\,^{\rm hyper}_{\rm visc} }{T} -\frac{ \eta_3\nabla^6\vec{A}\cdot\vec{j}}{\rho T} \, .
\end{equation}
We assume an ideal gas equation of state, with the density and entropy as the independent variables. Then, the fluid temperature is given by \citep{1959flme.book.....L}
\begin{equation}\label{eq:tem_def}
    T(\rho,s) = T_c \left(\frac{\rho}{\rho_c} \right) ^{\Gamma-1} \exp{[(s-s_c)/c_V]}\, ,
\end{equation}
where $s_c$ is a constant. Any variable with the subscript $c$, such as $\rho_c$ and $T_c$, refers to the initial value of the quantity at the center of the star. The pressure is
\begin{equation}\label{eq:pres_def}
    p(\rho,s)=(\mathcal{R}/\mu) \rho T(\rho,s) \, ,
\end{equation}
 where $\mathcal{R}$ is the universal gas constant and $\mu$ is the mean molecular weight. For all the simulations, we set $\mu=0.6$~g~mol$^{-1}$, as is typical for A stars \citep{2006A&A...450.1077B}, and the adiabatic index of a monatomic gas is
\begin{equation}\label{eq:adia_index}
    \Gamma\equiv\left(\frac{\partial {\rm ln}\, p}{\partial {\rm ln}\, \rho}\right)_s = \frac{c_p}{c_V}=\frac{5}{3}\, ,
\end{equation}
 with  $c_p$ and $c_V$ being the specific heats at constant pressure and volume, respectively.

Equations (\ref{eq:MHDEqs_a})-(\ref{eq:MHDEqs_d})  are numerically evolved with the {\sc Pencil Code}\footnote{ \texttt{https://github.com/pencil-code/}} \citep{2021JOSS....6.2807P},  a high-order finite-difference code for compressible hydrodynamic flows with magnetic fields. It  uses sixth-order centered spatial derivatives and a third-order Runge–Kutta time-stepping scheme. It is worth saying that the MHD equations are solved in terms of the vector potential, ensuring that the magnetic field remains divergence-free. The units used in the code are:
$$[x]=R_s, \quad  \quad [\rho]=\rho_c,\quad  \quad [t] = \left(G\rho_c\right)^{-1/2},  $$
$$[s]=c_p, \quad  \quad [B] =\sqrt{\mu_0G}\rho_cR_s, $$
with $G$ the gravitational constant, and $R_s$ the stellar radius. 

To reduce the numerical computation time, the gravitational potential is kept fixed along the simulation (Cowling approximation). In non-convective stars, the ratio between the fluid pressure and the pressure of the long-lived magnetic field is at least $\beta\equiv 2\mu_0 p/B^2 \sim  10^6$ (e.g., \citealt{2009A&A...499..557R}), so their non-magnetic, spherical hydrostatic equilibrium configuration will only be slightly perturbed by the presence of the magnetic field.

The parameters used for the simulations presented and discussed in this paper are summarized in Table~\ref{tab:models}.  We perform all the simulations in a cubic computational box of side $L_{\rm box} $ with the star at its  center. We use an equally-spaced Cartesian grid ($N_x=N_y=N_z\equiv N$, with $N_i$ the number of cells in the $i$-direction) with periodic boundary conditions. In Appendix~\ref{ap:test_sims}, we  test the accuracy of the code by comparing different box sizes and resolutions.

\begin{table*}
    \centering
    \addtolength{\tabcolsep}{-0.5pt}  
    \begin{tabular}{ccccccccccc}
    \hline
      Model   &   $\gamma$ &  $ \cfrac{\tau_s }{\tau_{\rm A,0}} $ &  $\cfrac{\tau_{\eta}}{\tau_{\rm A,0}} $ & $m$ & $r_0 $ & $\eta$-profile & $\eta_{\rm i}$ & $\nu$ &  $\nu_3$ & $\eta_3$ \\
       
        &    &    & & & $[R_s]$  &  & $[R_s^2/\tau_{\rm A,0}]$ & $[R_s^2/\tau_{\rm A,0}]$& $[R_s^6/\tau_{\rm A,0}]$ & $[R_s^6/\tau_{\rm A,0}]$  \\ \hline
      
        I & $4/3$   & $0.07$ & $2.41$ & $-4$ & $0.25$ & eq.~(\ref{eq:eta}) ($R_i=R_s$)&$0.0$ &$ 0.021 $ & $0.0$ & $0.0$ \\
        
        Ia &  $4/3$    & $0.07$ & $2.41$ & $-4$ & $0.25$ & eq.~(\ref{eq:eta_step}) &$-$  &$ 0.021 $ & $0.0$ & $0.0$ \\
        
        Ib &  $4/3$    & $0.07$ & $2.41$ & $-4$ & $0.25$ &  eq.~(\ref{eq:eta}) ($R_i=R_s$) &$-$  &$ 0.041 $ & $0.0$ & $0.0$ \\
        
        Ic &  $4/3$    & $0.07$ & $2.41$ & $-4$ & $0.25$ &  eq.~(\ref{eq:eta}) ($R_i=R_s$) &$-$  &$ 0.012 $ & $0.0$ & $0.0$ \\
         
        Id &  $4/3$    & $0.07$ & $2.41$ & $-4$ & $0.25$ &  eq.~(\ref{eq:eta}) ($R_i=R_s$) &$-$  &$ 0.0041 $ & $0.0$ & $0.0$ \\
        
        Ie &  $4/3$    & $0.07$ & $2.41$ & $-4$ & $0.25$ &  eq.~(\ref{eq:eta}) ($R_i=R_s$) &$-$  &$ 0.00041 $ & $0.0$ & $0.0$ \\
        
        If &  $4/3$   & $0.07$ & $2.41$ & $-4$ & $0.25$ & eq.~(\ref{eq:eta}) ($R_i=R_s$)&$4.1\times10^{-5}$ &$ 0.021 $ & $0.0$ & $0.0$ \\
        
        Ig &  $4/3$   & $0.07$ & $2.41$ & $-4$ & $0.25$ & eq.~(\ref{eq:eta}) ($R_i=R_s$)&$2.1\times10^{-4}$ &$ 0.021 $ & $0.0$ & $0.0$ \\
        
        Ih &  $4/3$    & $0.07$ & $2.41$ & $-4$ & $0.25$ & eq.~(\ref{eq:eta}) ($R_i=R_s$)&$4.1\times10^{-4}$ &$ 0.021 $ & $0.0$ & $0.0$ \\
        
        II & $4/3$   & $0.07$ & $2.41$ & $-4$ & $0.25$ & eq.~(\ref{eq:eta}) ($R_i=R_s$)&$0.0$ &$ 0.0 $ & $4.4\times 10^{-9}$ & $0.0$ \\
        
        III &  $4/3$    & $0.07$ & $2.41$ & $-4$ & $0.25$ & eq.~(\ref{eq:eta}) ($R_i=R_s$)& $0.0$ &$ 0.0 $ & $4.4\times 10^{-11}$ & $4.4\times 10^{-11}$ \\
        
        IV &  $4/3$     & $0.07$ & $2.41$ & $-4$ & $0.25$ & eq.~(\ref{eq:eta}) ($R_i=R_s$)&  $0.0$ & $ 0.0 $ & $2.2\times 10^{-10}$ & $2.2\times 10^{-10}$ \\
        
        V &  $4/3$   & $0.07$ & $2.41$ & $-4$ & $0.25$ & eq.~(\ref{eq:eta}) ($R_i=R_s$)& $0.0$ &$ 0.0 $ & $4.4\times 10^{-10}$ & $4.4\times 10^{-10}$ \\
        
        Va & $4/3$   & $0.07$ & $2.41$ & $-4$ & $0.25$ & eq.~(\ref{eq:eta_step}) & $-$ & $ 0.0 $ & $4.4\times 10^{-10}$ & $4.4\times 10^{-10}$ \\
        
        VI &  $4/3$    & $0.07$ & $2.41$ & $-4$ & $0.25$ & eq.~(\ref{eq:eta}) ($R_i=R_s$)& $0.0$ &$ 0.0 $ & $2.2\times 10^{-9}$ & $2.2\times 10^{-9}$ \\
        
        VII &  $4/3$   & $0.07$ & $2.41$ & $-1.5$ & $0.25$ & eq.~(\ref{eq:eta})  ($R_i=R_s$) & $0.0$ & $ 0.021 $ & $0.0$ & $0.0$ \\
        
        VIII &  $4/3$   & $0.07$ & $2.41$ & $-2.0$ & $0.25$ & eq.~(\ref{eq:eta})  ($R_i=R_s$)& $0.0$& $ 0.021 $ & $0.0$ & $0.0$ \\
        
        IX & $4/3$  & $0.07$ & $2.41$ & $-3.0$ & $0.25$ & eq.~(\ref{eq:eta})  ($R_i=R_s$)&$0.0$ & $ 0.021 $ & $0.0$ & $0.0$ \\
        
        X & $4/3$   & $0.07$ & $2.41$ & $-4$ & $0.35$ & eq.~(\ref{eq:eta})  ($R_i=R_s$)&$0.0$ &$ 0.021 $ & $0.0$ & $0.0$ \\
        
        Xa &  $4/3$   & $0.07$ & $2.41$ & $-4$ & $0.35$ & eq.~(\ref{eq:eta})  ($R_i=R_s$)&$0.0$ &$ 0.0 $ & $4.4\times 10^{-10}$ & $4.4\times 10^{-10}$ \\

        XI &  $4/3$  & $0.07$ & $2.41$ & $-4$ & $0.5$ & eq.~(\ref{eq:eta})  ($R_i=R_s$)&$0.0$ &$ 0.021 $ & $0.0$ & $0.0$ \\
        
        XIa &  $4/3$   & $0.07$ & $2.41$ & $-4$ & $0.5$ & eq.~(\ref{eq:eta})  ($R_i=R_s$)&$0.0$ &$ 0.0 $ & $4.4\times 10^{-10}$ & $4.4\times 10^{-10}$ \\
        
        XII &  $4/3$  & $0.07$ & $2.41$ & $-4$ & $0.7$ & eq.~(\ref{eq:eta})  ($R_i=R_s$)&$0.0$ &$ 0.021 $ & $0.0$ & $0.0$ \\
        
        XIIa &  $4/3$   & $0.07$ & $2.41$ & $-4$ & $0.7$ & eq.~(\ref{eq:eta})  ($R_i=R_s$)&$0.0$ &$ 0.0 $ & $4.4\times 10^{-10}$ & $4.4\times 10^{-10}$ \\
        
        XIII & $5/3$  & $0.09$ & $3.84$ & $-4$ & $0.25$ & eq.~(\ref{eq:eta})  ($R_i=0.9R_s$)&$0.0$ &$ 0.021 $ & $0.0$ & $0.0$ \\
        
        XIIIa &  $5/3$  & $0.09$ & $3.84$ & $-4$ & $0.25$ & eq.~(\ref{eq:eta})  ($R_i=0.9R_s$)&$0.0$ &$ 0.0 $ & $4.4\times 10^{-10}$ & $4.4\times 10^{-10}$ \\
    \hline
        
    \end{tabular}
    \caption{Parameters for the simulations done in this paper.  All the simulations are performed in a computational box of side $L_{\rm box}= 4.5 \,R_s$ and at a resolution of $128^3$. The adiabatic index is $\Gamma=5/3$, $E_{\rm mag}/|E_{\rm grav}|=0.0012$, and $\eta_{\rm ext}=0.414~R_s^2/\tau_{A,0}$. }
   \label{tab:models}
\end{table*}
%

\subsection{Initial set-up of the non-magnetic stellar model}\label{sec:int_setup}

Following \cite{2006A&A...450.1077B}, inside the star we adopt, as an initial condition, a polytropic relation between the gas pressure and density: $p=K \rho^{1+1/n}$, where $K$ is constant and $n$ is the conventional polytropic index. Thus,
\begin{equation}\label{eq:poly_index}
    \gamma = \frac{d\, {\rm ln}\, p}{d\, {\rm ln}\, \rho} = 1 + \frac{1}{n}.
\end{equation}
We use a polytropic index $n=3$ ($\gamma=4/3<\Gamma$) as an approximation for a radiative (stably stratified) star, while $n=1.5$ ($\gamma=5/3=\Gamma$) is used to model a convective (barotropic) star. 
Outside the star, there is a transition zone  that connects the star with a uniform temperature atmosphere (discussed in more detail in section \ref{sec:atm}). The temperature in this region has been increased for numerical reasons; otherwise, the density would become too small, making the Alfv\'en speed too high for numerical computation. 

Figure~\ref{fig:init_cond} shows the radial profile of the initial density, entropy, and temperature for $n=3$ and $n=3/2$. For a more detailed description on how  these profiles were built, see Appendix~\ref{ap:star}.

\subsection{Random initial magnetic field}\label{sec:randomField}

All the simulations start with a random magnetic field concentrated in the central region of the star. Following \cite{2015MNRAS.447.1213M}, each component of the magnetic vector potential is built in wavenumber space by assigning a random amplitude to each wave vector, $\vec{k}=\{k_x, k_y, k_z\}$ (the discrete values for the $i$-component of the wave vector are: $k^{(j)}_i = 2\pi j/L_{\rm box}$ for $j = -N/2, \cdots , N/2$; to guarantee periodic boundary conditions), such that:
\begin{equation}\label{eq:fourie_distr}
A_i(\vec{k}) = \left({\rm cos}\,\varphi_1 + i\,{\rm sin}\,\varphi_2 \right) |\vec{k}|^{m} \, , \; {\rm for}\quad \tilde{k}_{\rm 0}< |\vec{k}| <\tilde{k}_{\rm 1}
\end{equation}
where $\varphi_1$ and $\varphi_2$ are random numbers between $0$ and $2\pi$, $m (<0)$ a model parameter, $\tilde{k}_{\rm 0} = 2 \pi/R_s $, and $\tilde{k}_{\rm 1} = \pi N / (2L_{\rm box})$. Then, an inverse Fourier transformation is performed. In order to confine the initial magnetic field to the central region of the star, the resulting magnetic vector potential is multiplied by $\exp(-r^2/r_0^2)$, with $r_0$ a model parameter. Finally, the magnetic field amplitude is scaled in order to obtain a certain value of the Alfv\'en travel time (see below).

\subsection{Magnetic diffusivity in the atmosphere}
\label{sec:atm}

The star's atmosphere is taken to have a low electrical conductivity in order to make the external magnetic field relax to a potential field. So, we use a low  magnetic diffusivity  inside the star and a higher constant value outside it, with a transition zone connecting the two regions, i.e.,
\begin{equation}\label{eq:eta}
    \eta(r)= \left\{ 
     \begin{aligned}
&\eta_{\rm i}  \quad &
\mathrm{for}&\quad r < R_i, \\
&\frac{\eta_{\rm ext} -\eta_{\rm i}}{\Delta_r} \left(r-R_i\right)+\eta_{\rm i} 
\quad &
\mathrm{for}&\quad   R_i\leq r < R_i + \Delta_r, \\
  &\eta_{\rm ext} \quad &
\mathrm{for}&\quad r \geq R_i+\Delta_r.
\end{aligned} 
\right.
\end{equation}
with $\eta_{\rm i}$ and $\eta_{\rm ext}=0.414~R_s^2/\tau_{A,0}$, as the magnetic diffusivity inside the star and in the atmosphere, respectively ($\eta_{\rm i}<\eta_{\rm ext}$); and $
\Delta_r=0.3R_s$, the width of the transition zone. Unless otherwise specified, we use this profile in all the simulations. Since this diffusivity is not realistic, we neglect the corresponding heating term in equation~(\ref{eq:MHDEqs_d}), as was also done by \citet{2006A&A...450.1077B}. In fact, the magnetic energy dissipated by the magnetic diffusivity outside the star is $\sim 10^{-10}$ times the total magnetic energy change in the simulations. 

\section{Characterization of the magnetic field configurations}\label{sec:definitions}
To quantitatively evaluate whether a magnetic field configuration has an axially symmetric geometry, we introduce the asymmetry parameter, 
\begin{equation}\label{eq:axial_param}
\mathcal{A}
=\min_{\theta,\phi}\frac{\int_{\rm star} |\vec{B}-\vec{B}^{\rm axial}_{\theta,\phi} |^2 \,dV }{\int_{\rm star} |\vec{B} |^2\, dV } ,
\end{equation}
where $\vec{B}^{\rm axial}_{\theta,\phi}$ is built by taking the azimuthal average of the magnetic field components in spherical coordinates around a certain axis oriented in the direction given by the angles $\theta$ and $\phi$.  

The angles $\theta=\theta_0$ and $\phi=\phi_0$ that minimize $\mathcal{A}$ define the magnetic axis of the configuration, $\vec{M}$. For an axisymmetric magnetic field, $\mathcal{A}=0$.  Using a spherical coordinate system $(r,\theta,\phi)$ aligned with $\vec{M}$, it is possible to decompose the magnetic field into a ``toroidal'' component, $\vec{B}_{\rm tor} = (\vec{B}\cdot \hat{\phi}) \hat{\phi}$, with $\hat{\phi}$ the azimuthal unit vector, and a ``poloidal'' component, $\vec{B}_{\rm pol}= \vec{B} - \vec{B}_{\rm tor}$. If the field is axially symmetric, these are the usually defined toroidal and poloidal components, each of which is divergence-free, but this is not the case if the axial symmetry is broken.

Another quantity used to characterize the magnetic field configurations is the average wavenumber of the magnetic energy, defined as 
\begin{equation}\label{eq:k_mean}
\langle k \rangle =  \left( \sum_k 
k E(k)  \right)  \left( \sum_k
E(k) \right)^{-1} ,
 \end{equation}
with $k$ the wavenumber, which takes values from 0 to $\pi N/ L_{\rm box}$ in intervals $\Delta k= 2\pi / L_{\rm box}$, and $E(k)$ corresponds to the magnetic field energy spectrum, defined as
\begin{equation}\label{eq:Spectrum}
E( k ) =  \sum_{ |\vec{k}|-\Delta k/2<|\vec{k'}|<|\vec{k}|+\Delta k/2 } \frac{1}{2} \vec{B}_{\vec{k'}} \cdot \vec{B}^{*}_{\vec{k'}}\, ,
 \end{equation}
where 
\begin{equation}\label{eq:FT_Bfield}
    \vec{B}_{\vec{k}} = \sum_{l,m,n} \vec{B}_{l,m,n} e^{-i\vec{k}\cdot\vec{r}_{l,m,n}}\, 
\end{equation}
is the discrete Fourier transform of $\vec{B}$, and the asterisk denotes its complex conjugate. The subscript $A_{l,m,n}$ denotes the value of $A$ at the grid point $r_{l,m,n}=(x_l,y_m,z_n)$, and  the sum is made over all of them.  

The magnetic, kinetic,  internal, and gravitational energy  are calculated as
\begin{align}\label{eq:emag}
E_{\rm mag} & = \cfrac{1}{2\mu_0} \int_{\rm box} |\vec{B}|^2 \,dV, \\
E_{\rm kin} & = \cfrac{1}{2}\int_{\rm box}  \rho|\vec{v}|^2 \,dV, \label{eq:ekin} \\
E_{\rm int} & = \int_{\rm box}  \rho c_v T \,dV =\int_{\rm box}  \cfrac{p}{\Gamma-1} \,dV, \label{eq:Eint}\\
E_{\rm grav} &= \cfrac{1}{2}\int_{\rm box} \rho\Phi \,dV,\label{eq:egrav}
\end{align}
respectively. The total energy of the simulation box (the sum of these four terms) should be conserved up to the discretization error\footnote{See \texttt{http://pencil-code.nordita.org/doc/manual.pdf}}. In fact, the non-conservation of this quantity can be taken as a sign of relevant numerical errors in the simulation. 

We also define the Alfv\'en and the sound crossing time scales as
\begin{equation}
\tau_{\rm A} \equiv \frac{R_s\sqrt{\mu_0\rho_{\rm rms}}}{B_{\rm rms}} \qquad {\rm and} \qquad \tau_{\rm s} \equiv \frac{R_s}{c_{\rm s, rms}}\, ,
\end{equation}
respectively, where $c_s=\sqrt{\Gamma p/\rho}$ is the sound speed and the notation $A_{\rm rms}$ refers to the root-mean-square of the quantity $A$ over the volume of the star ($r\leq R_s$).

\section{Physical damping mechanism}\label{sec:PhyDampMch}

When we add a random magnetic field to the initially non-magnetic star in hydrostatic equilibrium, it will move the fluid and eventually, once the motion is damped, it may relax to a stable hydromagnetic equilibrium state. Possible damping mechanisms include viscosity and magnetic diffusivity. In a main-sequence star, the viscosity coefficient is given by \citep{Braginskii1965}
\begin{equation}\label{eq:nu_def}
\nu \sim  \frac{(\kappa_B T)^{5/2} }{m_i^{1/2} n_i (Z_ie)^4}\sim 10^{3} \left(\frac{T}{5\times 10^6\, \mathrm{K}}\right)^{5/2} \left(\frac{10^{23} \mathrm{cm}^{-3}}{n_i} \right)\, \mathrm{cm}^{2} \, \mathrm{s}^{-1} \, ,
\end{equation}
while the magnetic diffusivity is \citep{1953PhRv...89..977S}
\begin{equation}\label{eq:eta_def}
\eta \sim  \frac{\sqrt{m_e}c^2Z_ie^2}{(\kappa_B T)^{3/2}} \sim 300  \left(\frac{T}{5\times 10^6\, \mathrm{K}}\right)^{-3/2} \, \mathrm{cm}^{2} \, \mathrm{s}^{-1}\, , 
\end{equation}
where $m_i$, $Z_i$, and $n_i$ are the ion mass, charge number, and number density, respectively. Thus, the viscous and magnetic diffusion timescales can be estimated as 
\begin{equation}\label{eq:tau_nu}
    \tau_\nu =\frac{\lambda_B^2}{\nu}\sim 10^{11}\left(\frac{\lambda_B}{R_s}\right)^2\left(\frac{5\times 10^6\, \mathrm{K}}{T}\right)^{5/2} \left(\frac{n_i}{10^{23} \mathrm{cm}^{-3}} \right)\, \mathrm{yr}\, ,
\end{equation}
and
\begin{equation}\label{eq:tau_eta}
    \tau_\eta =\frac{\lambda_B^2}{\eta}\sim 10^{11}\left(\frac{\lambda_B}{R_s}\right)^2\left(\frac{5\times 10^6\, \mathrm{K}}{T}\right)^{-3/2} ~{\rm yr}\, ,
\end{equation}
where $\lambda_B$ is a characteristic length scale of the magnetic field. Both are too long for pure diffusion (of momentum or magnetic flux) to dissipate the kinetic or magnetic energy within the star's lifetime.  

It has been suggested that  the mechanism that allows the star to reach an equilibrium is phase mixing \citep{1999A&A...349..189S}. Due to the non-uniform Alfv\'en speed, waves oscillate out of phase with each other and very small-scale velocity and magnetic field gradients are built up, causing rapid damping of these waves by the dissipative processes mentioned above. The characteristic timescale of phase mixing is $\tau_{\rm ph}=\lambda_{\rm ph}/v_{A}$, where $\lambda_{\rm ph}$ is a damping length given by \cite{1983A&A...117..220H}
\begin{equation}\label{lph_def}
\lambda_{\rm ph}=\left[\frac{3}{(\nu+\eta) 2\pi^2 } \left( \frac{v_A}{|\vec{\nabla}_{\perp} v_A|}\right)^2v_A\lambda_B^2\right]^{1/3},
\end{equation}
$v_A=|\vec{B}|/\sqrt{\mu_0 \rho}$ is the Alfv\'en speed, and $\vec{\nabla}_{\perp}$ is the gradient in the direction perpendicular to the magnetic field. Thus,
\begin{equation}
\tau_{\rm ph}\sim 10^4\left[\left(\frac{\tau_{\nu,\eta}}{10^{11}~\mathrm{yrs}}\right)\left(\frac{\lambda_C}{0.45R_s}\right)^2\left(\frac{10^3~G}{B}\right)^{2}\left(\frac{n_i}{10^{23}\mathrm{cm}^{-3}}\right)\right]^{1/3}~\rm{yr}\, ,
\end{equation}
where $\lambda_C\equiv v_A/|\vec{\nabla}_{\perp} v_A|$. Phase mixing damps the motion inside the star and leads to a hydromagnetic equilibrium in a timescale much shorter than the star's lifetime.

Similar estimates can be made for degenerate stars. In the case of white dwarfs, the typical dissipation time scale is longer than the star’s lifetime \citep{1999A&A...346..345P,2018ASSL..457..455S}, while for the case of neutron star cores, the viscous time-scale is much shorter than the magnetic diffusivity timescale  \citep{1969Natur.224..674B,2018ASSL..457..455S}. In these cases, the phase-mixing timescale is also shorter than the viscous or magnetic diffusivity time scales.

Due to the impossibility of resolving the phase-mixing process in the simulations, we will use high values for the viscosity and hyper-viscosity to model the relaxation of the magnetic field to an equilibrium configuration, because the (hyper-)viscous force, like phase mixing, stops acting when there is no motion inside the star, i.~e., when a stable equilibrium has been reached. We also explore the effect on the magnetic field evolution of introducing  magnetic diffusivity and hyper-diffusivity, as done in previous work. In the following sections, we present our results.

\section{Simulation results}\label{sec:sim_results}

\subsection{Effects of viscosity and hyper-viscosity}\label{sec:sim_visc}

%
\begin{figure}
    \centering
    \includegraphics[width=0.95\columnwidth]{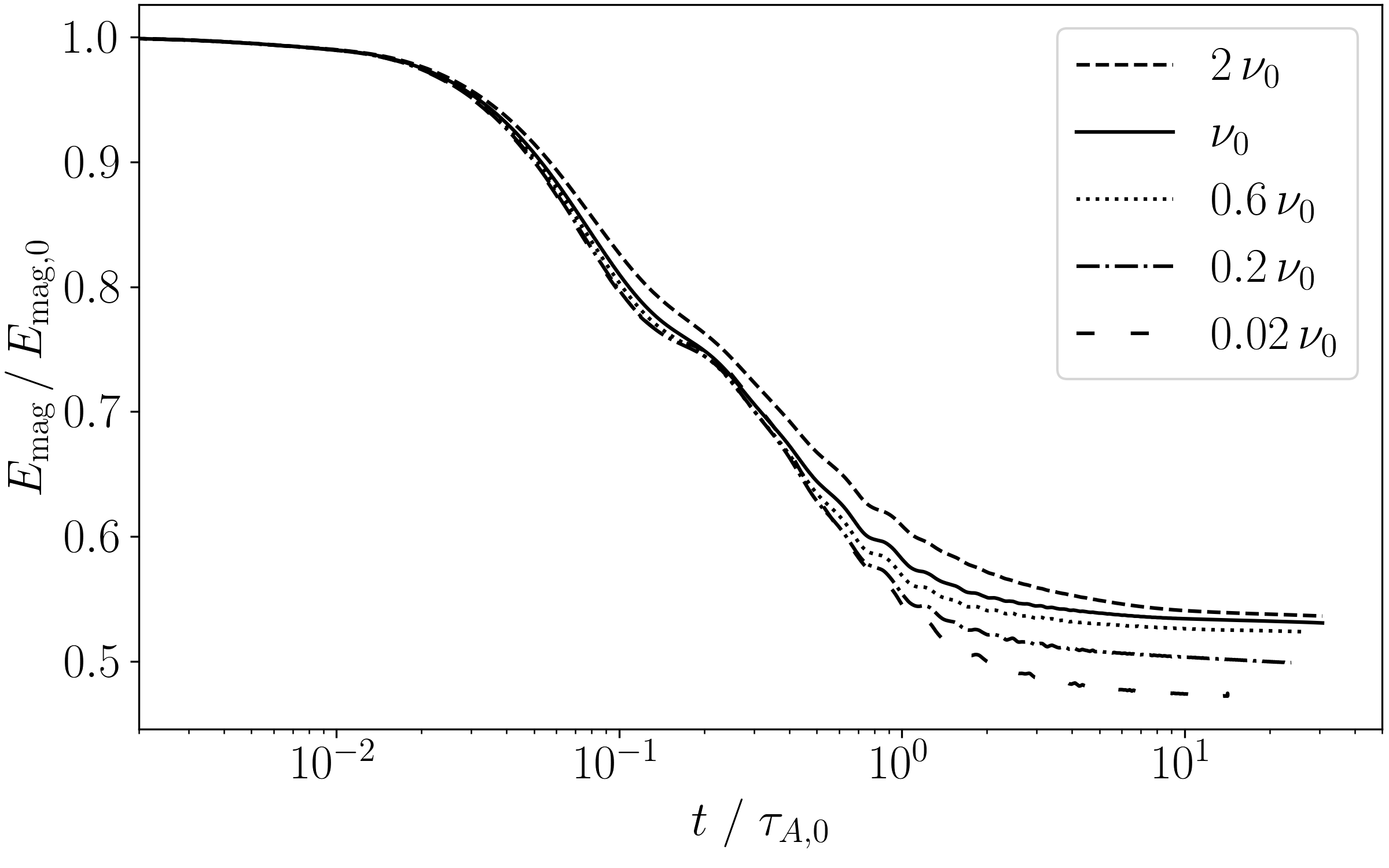}
    \caption{  Magnetic energy evolution for four different values of the viscosity coefficient (Models I, Ib, Ic, Id and Ie from Table~\ref{tab:models} with $\nu_0=0.021 R_s^2/\tau_{A,0}$).
    }
    \label{fig:Emag_visc}
\end{figure}
\begin{figure}
    \centering
    \includegraphics[width=0.95\columnwidth]{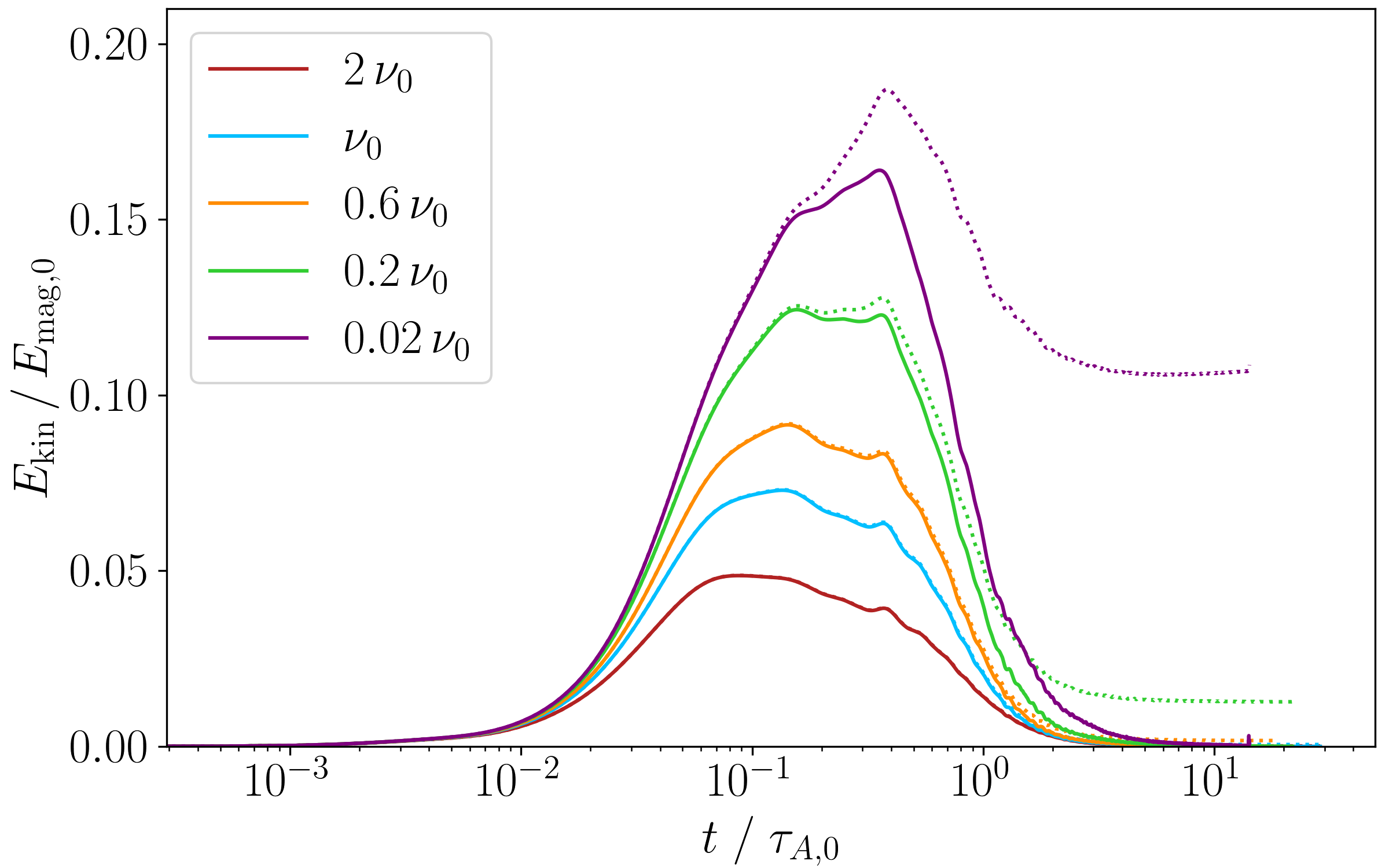}
    \caption{  Time evolution of the kinetic energy  for Models Ib, I, Ic, and Id from Table~\ref{tab:models} (with regular viscosity and $\nu_0=0.021 R_s^2/\tau_{A,0}$) . Solid lines correspond to the kinetic energy given by equation (\ref{eq:ekin}) while dotted lines to the time integration of equation (\ref{eq:ekindot_a}).
    }
    \label{fig:Ekin_visc}
\end{figure}
\begin{figure}
    \centering
    \includegraphics[width=0.95\columnwidth]{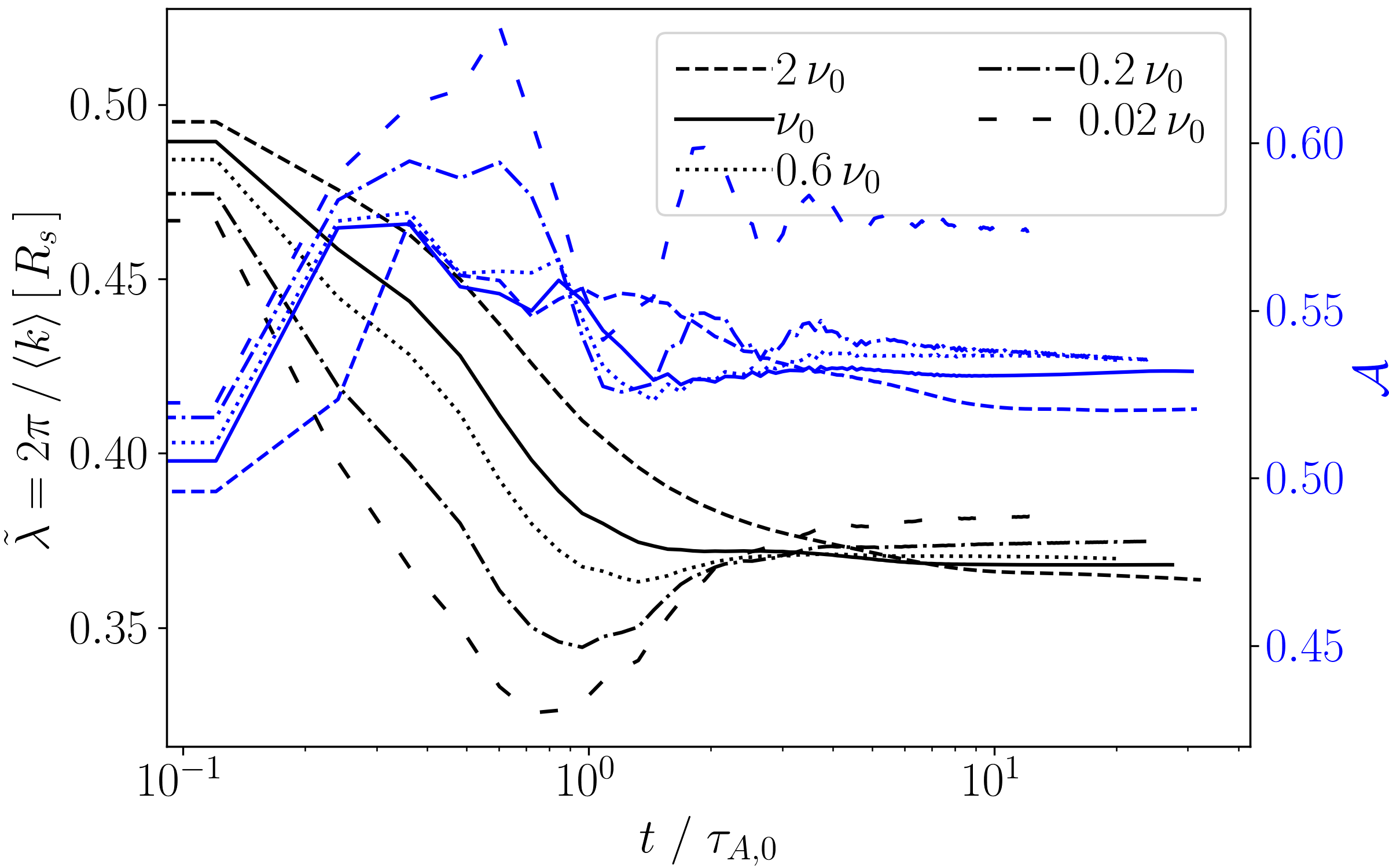}
    \caption{Time evolution of the characteristic wavelengths (defined in terms of the average wavenumber, equation~\ref{eq:k_mean}) of the magnetic energy spectrum (black curves and left vertical axis) and the asymmetry parameter (blue curves and right vertical axis)  for the same  Models Ib, I, Ic, Id and Ie from Table~\ref{tab:models} with $\nu_0=0.021 R_s^2/\tau_{A,0}$.
    }
    \label{fig:kmean_visc}
\end{figure}
\begin{figure}
    \centering
    \includegraphics[width=\columnwidth]{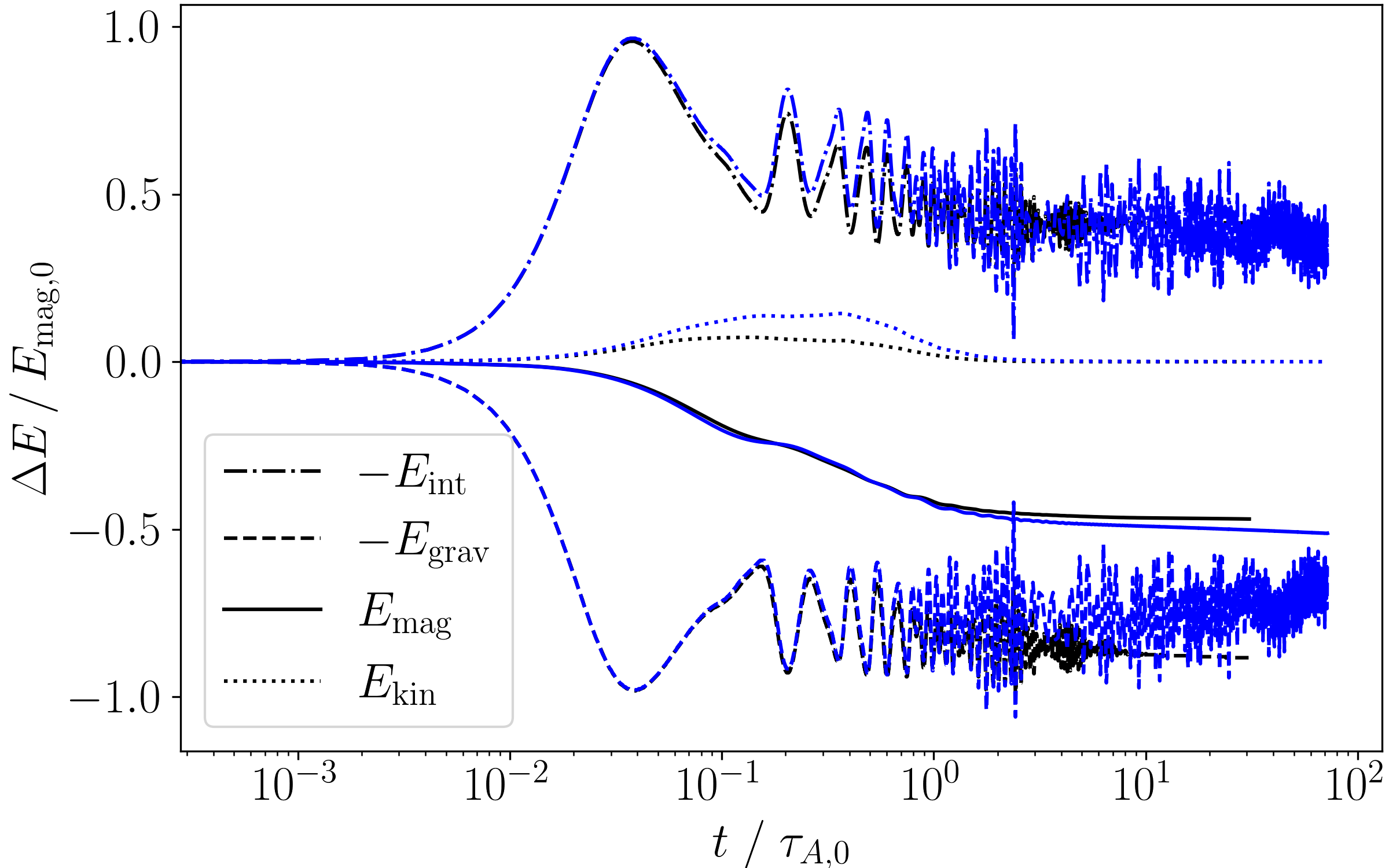}
    \caption{Time evolution of the cumulative change in internal energy, $E_{\rm int}$, gravitational energy, $E_{\rm grav}$ (both with their sign changed for presentation purposes), magnetic energy, $E_{\rm mag}$, and kinetic energy, $E_{\rm kin}$, for Model I (in black), with ordinary viscosity, and Model II (in blue), with hyper-viscosity, of Table~\ref{tab:models}. }
    \label{fig:energies}
\end{figure}
\begin{figure*}
    \centering
    \includegraphics[width=0.99\textwidth]{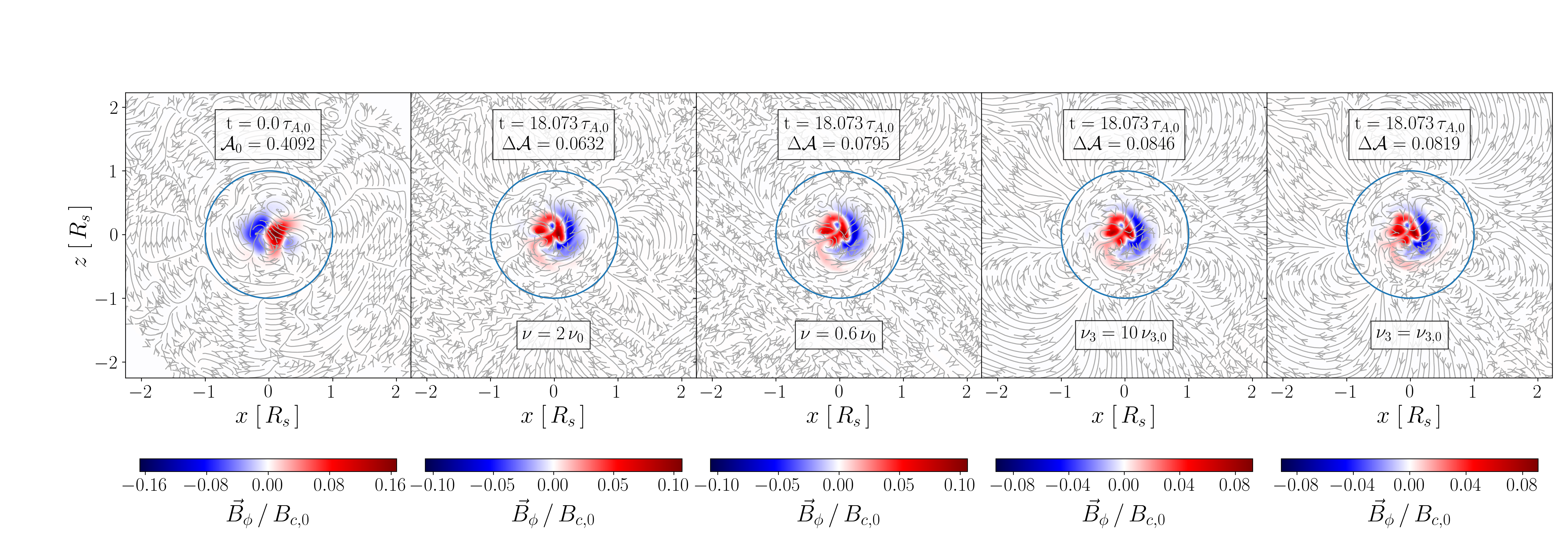}
    
    \caption{Snapshots of the magnetic field for Models Ib (with $\nu=2\nu_0$), Ic (with $\nu=0.6\nu_0$) and Model II (with $\nu_3=\nu_{3,0}$) from Table~\ref{tab:models} and Model AIa from Table~\ref{tab:models_appendix}
    at the indicated times, showing the magnetic field lines of the poloidal component and a color scale corresponding to the strength and direction of the toroidal component, in a meridional cut through the star. The vertical axis is aligned with the instantaneous magnetic
    axis $\vec{M}$,  and  the blue circle represents the star's surface. The initial value of the asymmetry parameter, $
\mathcal{A}$, is specified at the top of the left panel, while its change with respect to the initial state, $\mathcal{A}_0$,  is given at the top of the other panels. }
    \label{fig:Bfield_snap}
\end{figure*}

We start studying cases in which ordinary viscosity or hyper-viscosity provide the only explicit dissipation inside the star, with the magnetic diffusivity and hyper-diffusivity inside the star set to zero. As said before, we believe that this is the setup 
that best mimics the effects of phase mixing, since the kinetic energy will be dissipated by the (hyper-)viscosity, and, once the motion is damped inside the star, it leaves a magnetic field configuration in equilibrium with the pressure and gravitational forces, with no further evolution. 

In order to probe the dependence of the magnetic field evolution on  the viscous dissipation, we  run several    simulations of stably stratified stars (polytropic index $n=3$) that only differ in the value of the (ordinary) viscosity coefficient $\nu$. As seen in Figure~\ref{fig:Emag_visc}, the magnetic energy decreases more slowly for larger viscosities, as might be expected if the motion induced by the Lorentz force is mostly opposed by the viscous force. On the other hand, the total decrease of the magnetic energy is larger for smaller values of $\nu$, which is probably due to numerical dissipation present in the simulations, as we will now show.

To assess the importance of the numerical dissipation in these simulations, Figure~\ref{fig:Ekin_visc} compares the evolution of the kinetic energy calculated in two ways: One by directly doing the volume integration of equation~(\ref{eq:ekin}) at each moment of time, and the other one by numerically integrating the time derivative of the kinetic energy, which can be derived from equations ~(\ref{eq:MHDEqs_a}) and (\ref{eq:MHDEqs_b})
 as\footnote{For our periodic box, the surface term vanishes identically.}
\begin{align}\label{eq:ekindot_a}
    \dot{E}_{\rm kin} &=  \frac{1}{2}\int_{\rm box} \frac{\partial \rho |\vec{u}|^2}{\partial t} dV = - \int_S  \left(\frac{\rho |\vec{u}|^2 }{2}\right) \vec{u}\cdot d\vec{S} + \\
     &  \int_{\rm box} -\vec{u}\cdot \left( \vec{\nabla}p +\rho\vec{\nabla}\phi -  \vec{j}\times\vec{B} 
     - \rho\vec{f}_{\rm visc}  - \rho\vec{f}_{\rm visc}^{\rm hyper}  \right)
     dV \, . \nonumber
\end{align}
For all these simulations, the directly calculated kinetic energy peaks at $t\sim 0.1\tau_{A,0}$ and later decreases to zero. The peak is higher for smaller values of $\nu$, confirming that the viscosity always plays an important role in the evolution, likely because the motions are driven by the random initial magnetic field and thus vary on small spatial scales. For the largest values of $\nu$, the two ways of calculating $E_{\rm kin}$ are essentially in agreement, whereas they become progressively more discrepant as $\nu$ decreases, signaling that the numerical dissipation becomes relatively more important, to the point of dominating the dissipation of the kinetic energy after the peak for the smallest viscosity considered. Thus, the most accurate simulations are those with $\nu\gtrsim 0.021 R_s^2/\tau_{A,0}$. However, we note that none of these simulations are astrophysically realistic in the sense that, in real stars, all dissipative effects are expected to be negligible on time scales $\lesssim\tau_{A,0}$, which is clearly not the case here, and which cannot be realized because of the importance acquired by the numerical dissipation as $\nu$ decreases. 

As seen in Figure~\ref{fig:kmean_visc}, numerical dissipation also affects the final magnetic configuration by making it slightly more axisymmetric and dominated by slightly larger wavelengths, while for the simulations with $\nu\gtrsim 0.021 R_s^2/\tau_{A,0}$ the evolution leads to equilibrium magnetic field configurations with very similar mean wavelengths and asymmetry parameters. A similar analysis for the cases where only hyper-viscosity is dissipating the kinetic energy gives us that for $\nu_3>2.2\times10^{-9} R_s^6/\tau_{A,0}$ the numerical dissipation can be neglected (see appendix~\ref{ap:test_sims}).

Now that we have determined the values of $\nu$ and $\nu_3$ for which numerical dissipation can be neglected, we can contrast the effects of viscosity and hyper-viscosity by comparing Model I and Model II in  Table~\ref{tab:models}. For Model I, the kinetic energy is dissipated by ordinary viscosity ($\nu=0.021~R_s^2/\tau_{A,0}$), while for Model II, this is done by hyper-viscosity ($\nu_3 = 4.4 \times 10^{-9} ~R_s^6/\tau_{\rm A,0}$). Figure~\ref{fig:energies} shows that the time evolution of the gravitational, magnetic, kinetic, and internal energy is nearly identical for both simulations at $t\lesssim\tau_{A,0}$, when dissipative effects are relatively unimportant, and still quite similar at later times, when the dissipative effects are dominant. From the beginning and over much of the evolution, we see coherent oscillations of the thermal and gravitational energies, with periods comparable to $\tau_s$, which can be interpreted as  oscillations of the star due to its initial departure from equilibrium. Due to their long wavelength, these are relatively unaffected by either kind of viscosity. By time $t\sim 0.2\tau_{A,0}$, part of the magnetic energy has been converted to kinetic energy, which is at the same time being dissipated by viscous effects on small spatial scales set by the initial random magnetic field, making the magnetic field relax to a stable equilibrium in a time scale that, for the rather large viscosity parameters used here, is comparable to the Alfv\'en time, after which the magnetic energy no longer decays.\footnote{For simulations of Model II without the heating term, $\mathcal{H}^{\rm hyper}$, in equation~(\ref{eq:MHDEqs_d}), the star does not reach a stationary state, but the magnetic energy keeps decaying monotonically.}  Figure~\ref{fig:Bfield_snap} shows that, for different values of the viscosity or hyper-viscosity, the final equilibrium state of the magnetic field in the inner part of the star, where the field is mostly localized, is nearly identical. In strong contrast with the simulations of \cite{2006A&A...450.1077B}, which include hyper-diffusion terms in equations (\ref{eq:MHDEqs_b})-(\ref{eq:MHDEqs_d}) and find roughly axisymmetric equilibrium states, the final configurations found here are by no means axisymmetric. In fact, the net change in the asymmetry parameter is very small (and, in fact, positive for the simulations shown here), implying that the stable configuration is roughly as asymmetric as the initial random field.

We can draw several conclusions from the simulations analyzed in this subsection. First, they do not need to be stabilized by the hyper-diffusion scheme considered in equations (\ref{eq:MHDEqs_a})-(\ref{eq:MHDEqs_d}), since they can run for many Alfv\'en timescales using just ordinary viscosity. Second, the presence of numerical dissipation effects forced us to use relatively high values of the ordinary viscosity, $\nu$, or hyper-viscosity, $\nu_3$, in order to obtain accurate (energy-conserving) results. 
Unfortunately, these high values yield damping times comparable to the Alfv\'en time, that is, much shorter than expected in real stars. For these values, the final magnetic equilibrium state is roughly as asymmetric as the random initial magnetic field, and it is nearly independent of the values used for $\nu$ and $\nu_3$, leading us to speculate that it represents the final state reached also for the much smaller values relevant for real stars. Of course, in this case the detailed form of the final state will depend on the initial state of the magnetic field, which is set by its largely unknown formation mechanism, but in principle it could be much less symmetric than those obtained by \cite{2006A&A...450.1077B}.

\subsection{Effects of the magnetic (hyper-)diffusivity inside the star}\label{sec:sim_diffusivity}

%
 \begin{figure}
    \centering
    
    \includegraphics[width=0.99\columnwidth]{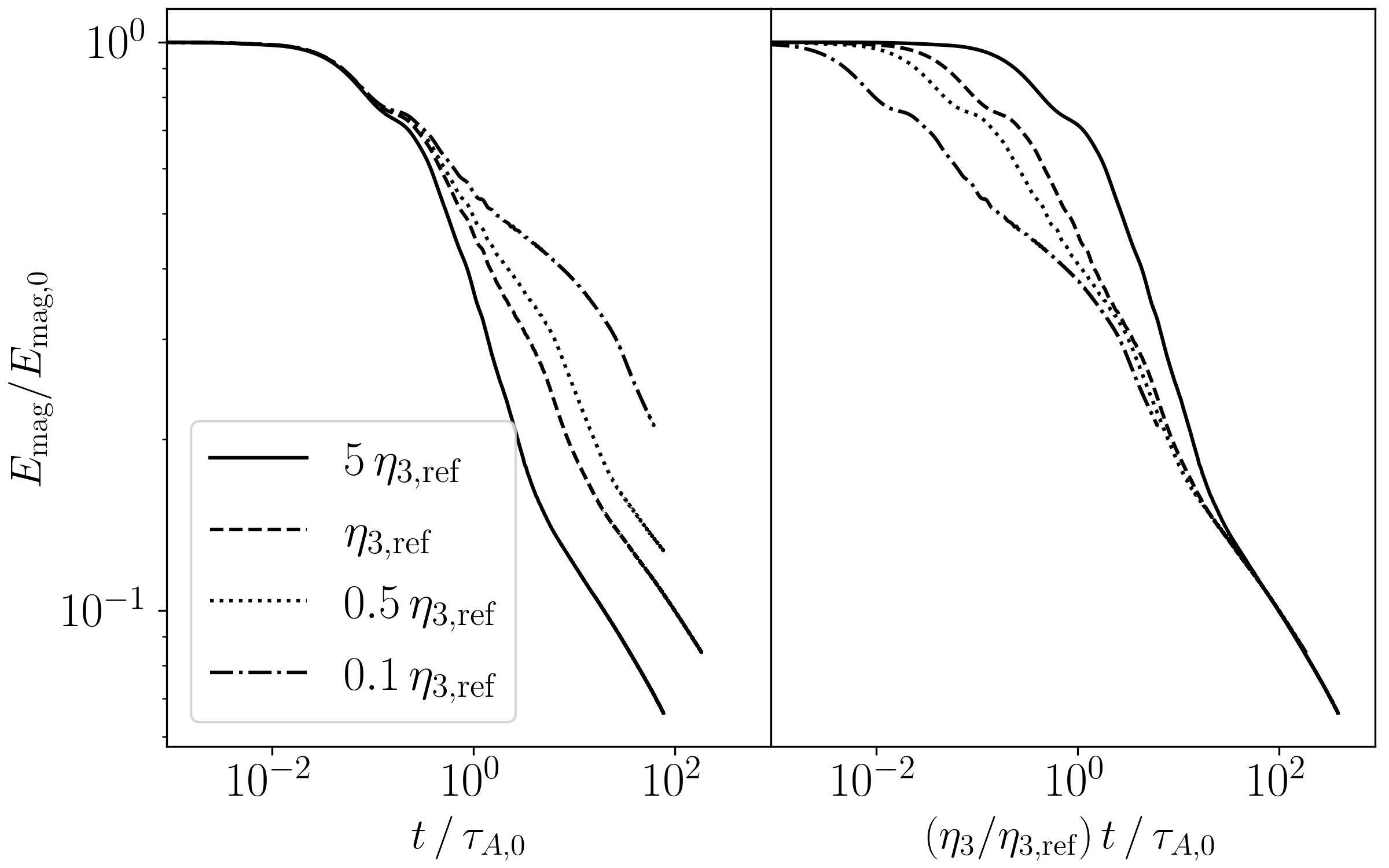}
    
    \caption{Magnetic energy evolution for four different values for the hyper-diffusion coefficients (always $\nu_3 = \eta_3$, expressed in terms of a reference value $\eta_{\rm 3, ref} = 4.4\times 10^{-10}\; R_s^6 /\tau_{\rm A,0})$ (Models III, IV, V, and VI of Table~\ref{tab:models}). Different normalizations of the time coordinate are used in the left and right panel. }
    \label{fig:Emag_hyper}
\end{figure}
\begin{figure}
    \centering
    
    \includegraphics[width=\columnwidth]{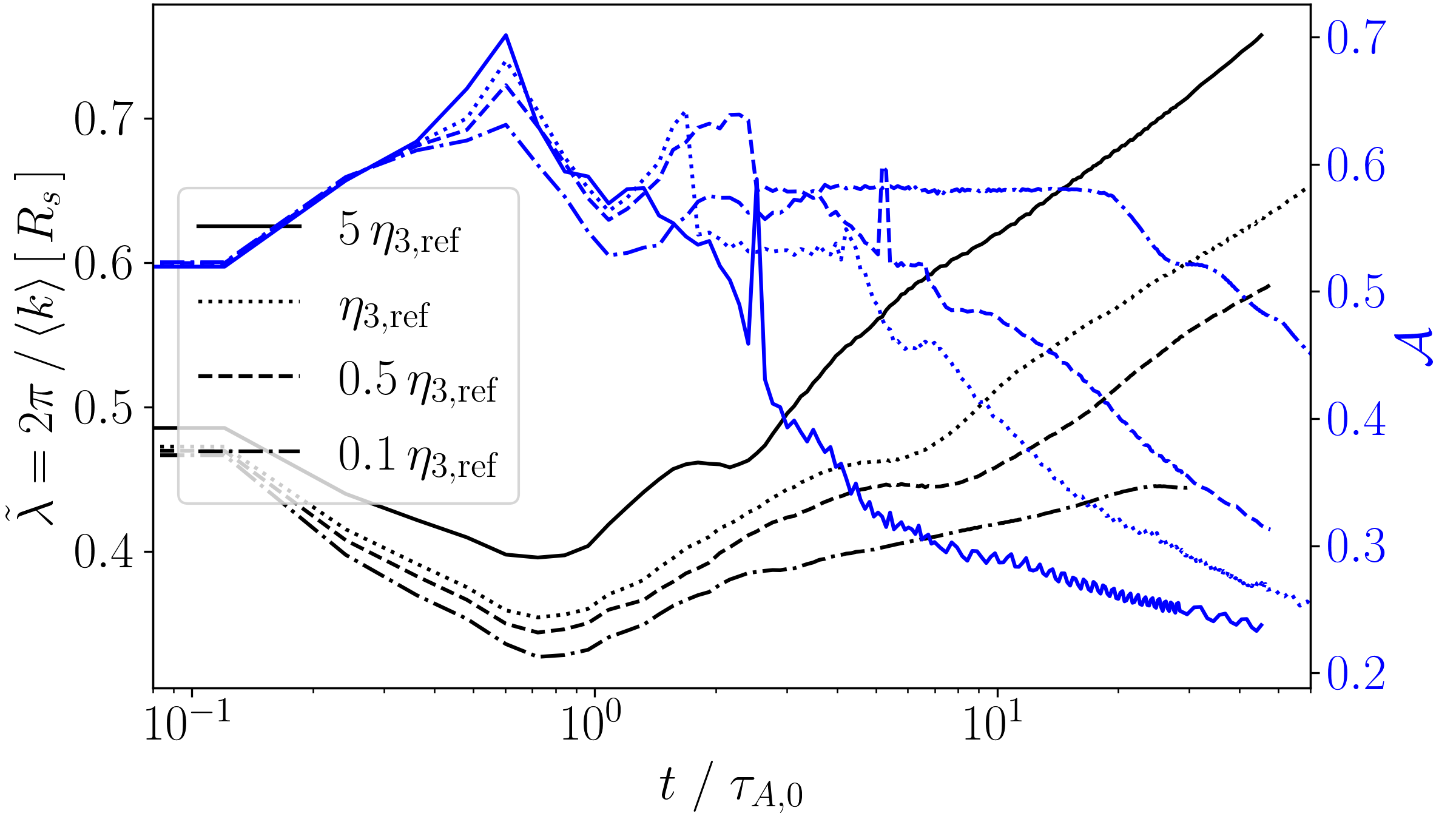}
    
    \caption{Same as Figure~\ref{fig:kmean_visc} but for the same models of  Fig.~\ref{fig:Emag_hyper} (Models  III, IV, V, and VI of Table~\ref{tab:models}), corresponding to different magnitudes of the hyper-diffusion coefficients, always with $\nu_3 = \eta_3$ given in terms of the reference value $\eta_{\rm 3, ref} = 4.4\times 10^{-10}\; R_s^6 /\tau_{\rm A,0})$.}
    \label{fig:Axial_hyper}
\end{figure}
\begin{figure}
    \centering
    \includegraphics[width=\columnwidth]{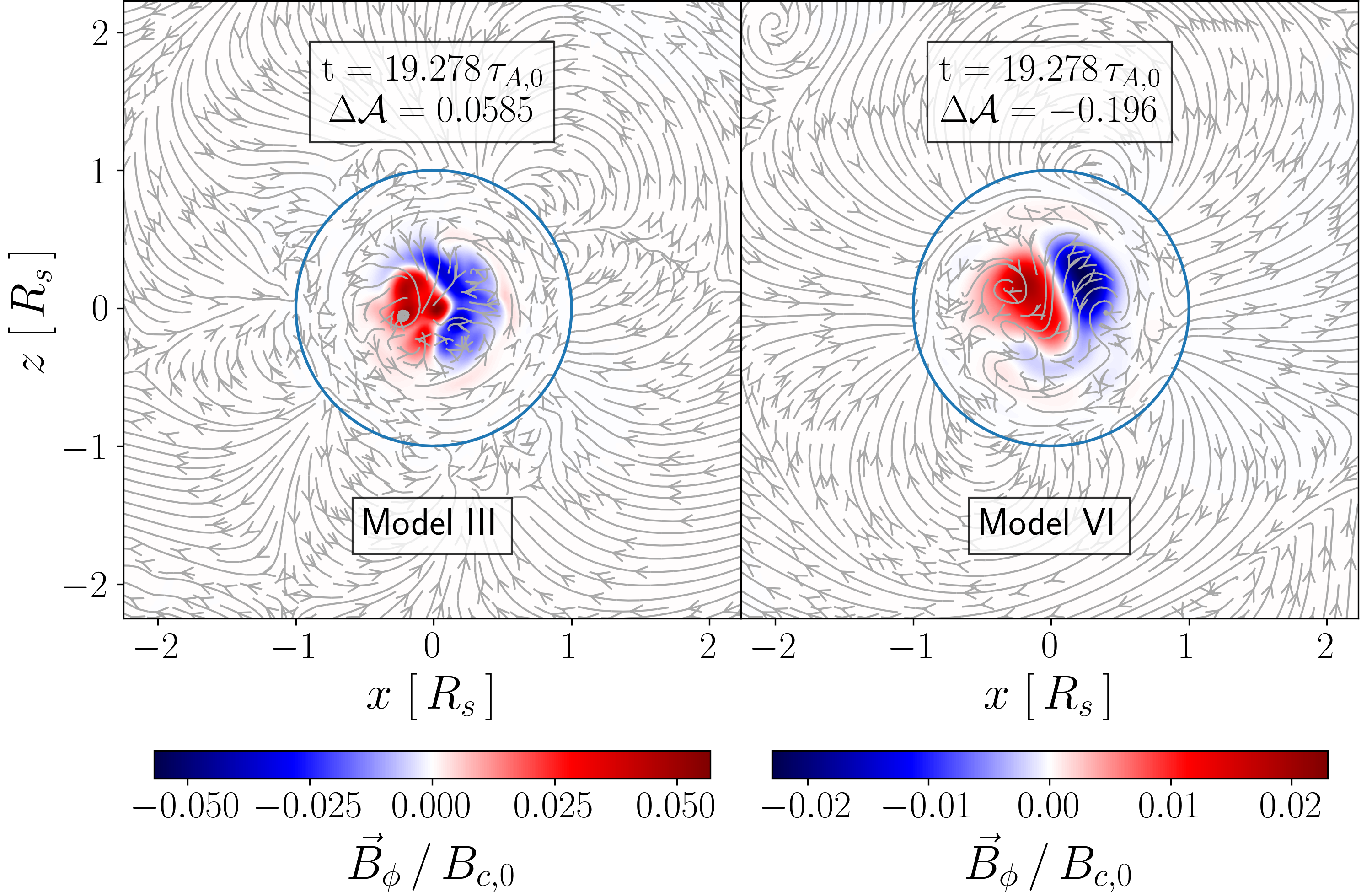}
    
    \caption{Same as Figure~\ref{fig:Bfield_snap} for Models III (with $\nu_3=\eta_3=0.2\eta_{\rm 3,ref}$)  and VI (with $\nu_3=\eta_3=5\eta_{\rm 3,ref}$)  in Table~\ref{tab:models} ($\eta_{\rm 3, ref} = 4.4\times 10^{-10}\; R_s^6 /\tau_{\rm A,0})$). }
    \label{fig:Bfield_hyper}
\end{figure}
\begin{figure}
    \centering
    \includegraphics[width=\columnwidth]{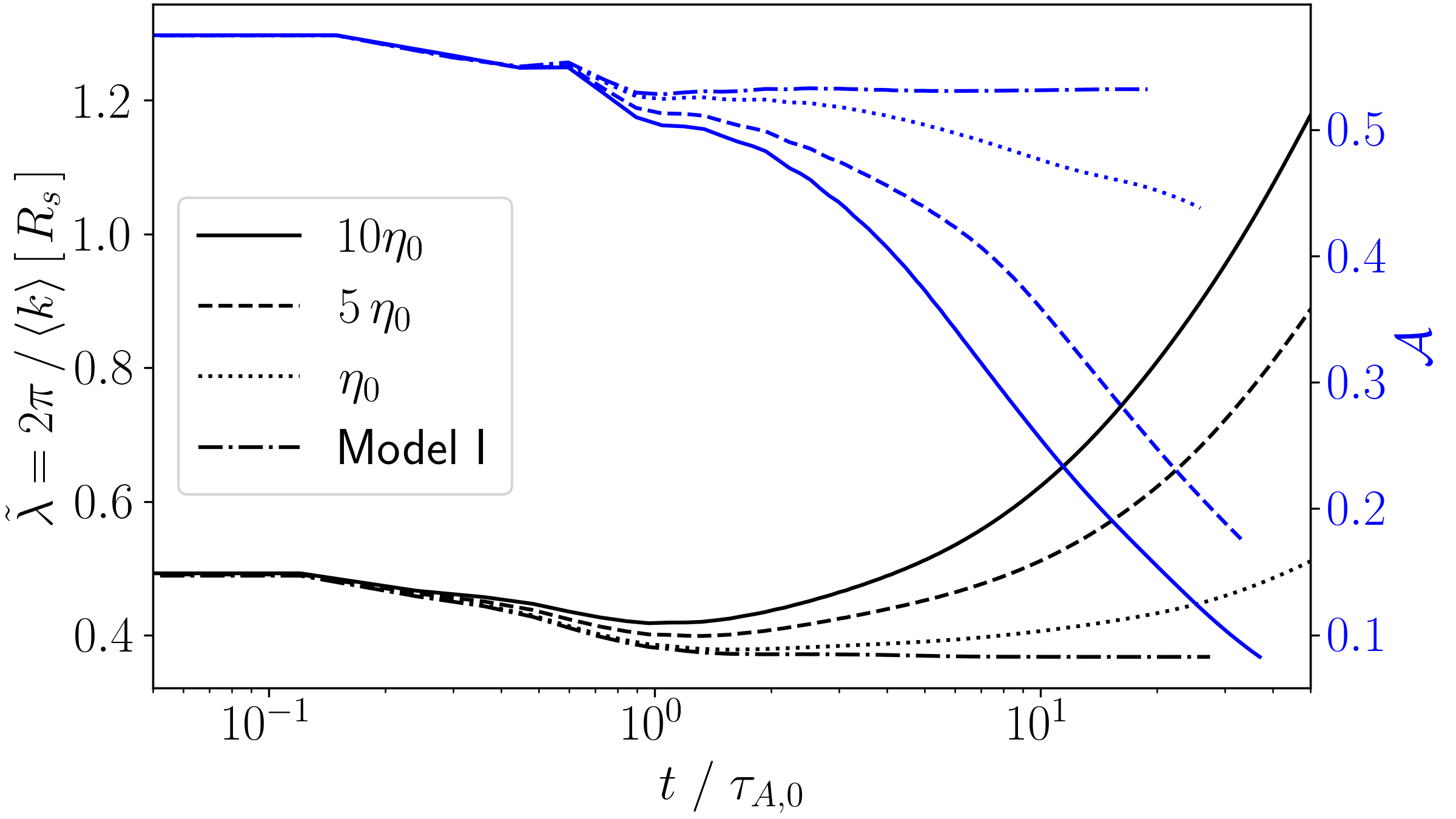}
    
    \caption{Same as Figure~\ref{fig:kmean_visc} for Models I($\eta_{\rm i}=0$), If ($\eta_{\rm i}=\eta_0$), Ig ($\eta_{\rm i}=5\eta_0$) and Ih ($\eta_{\rm i}=10\eta_0$) of Table~\ref{tab:models}, with  $\eta_0=4.1\times 10^{-5} R_s^2/\tau_{A,0}$. The dissipation mechanism of kinetic energy is ordinary viscosity. }
    \label{fig:Axial_mag}
\end{figure}
Although we have argued that the scenario that best reproduces the physical conditions inside real stars is the one with zero magnetic diffusivity, in this section, to facilitate comparison with the results of \citep{2006A&A...450.1077B}, we study how magnetic diffusivity and magnetic hyper-diffusivity change the evolution of the magnetic field.

First, in order to make our simulation closer to the one presented in \cite{2006A&A...450.1077B}, we introduce the effect of the  complete hyper-diffusion scheme,  including both the  hyper-viscosity force, $\vec{f}_{\rm visc}^{\rm hyper}$, and the magnetic hyper-diffusivity,  given by the last term of equation~(\ref{eq:MHDEqs_c}).  In principle, this scheme should provide a preferential dissipation of modes approaching the Nyquist frequency (high wavenumbers and  spatial scales comparable to the grid spacing) without affecting the modes with smaller wavenumbers. We note that the hyper-diffusion scheme implemented in our simulations is different to the one used in the {\rm Stagger code}, for which a direct comparison is not possible.

We run simulations with different values for the hyper-diffusion coefficients ($\nu_3$ and $\eta_3$, Models III to VI of Table~\ref{tab:models}) and compare the time evolution of the total magnetic energy in Figure~\ref{fig:Emag_hyper}.  In the left panel, this evolution is shown as a function of $t/\tau_{A,0}$, making it clear that, until $t\sim\tau_{A,0}$, the evolution is independent of these coefficients, i.~e., the hyper-diffusion is not important. 
In the right panel,  we plot the same curves as functions of $(\eta_3/\eta_{\rm 3,ref})(t/\tau_{A,0})$, in which case all of them converge  at $(\eta_3/\eta_{\rm 3,ref})(t/\tau_{A,0})\sim 30$, signaling that by this time  the hyper-diffusion has become dominant. Contrary to Models I and II, discussed in the previous section, the magnetic energy does not reach an asymptotic value, instead it is continuously dissipated by the hyper-diffusivity.

For the same models, Figure~\ref{fig:Axial_hyper} shows that, after a few Alfv\'en times, the characteristic wavelength of the magnetic field configuration tends to increase and its asymmetry parameter, $\mathcal{A}$, tends to decrease with time, i.~e., the hyper-diffusion scheme (mainly the magnetic hyper-diffusivity) makes the final magnetic equilibrium configuration more large-scale and axisymmetric than the random initial state, most likely because the hyper-diffusion preferentially dissipates the energy at the shortest wavelengths, leaving a magnetic field configuration dominated by the longest wavelengths. The effect is strongest for the larger values of the hyper-diffusion coefficient, for which the final stable magnetic field configuration looks more like the roughly axisymmetric twisted torus described in \cite{2006A&A...450.1077B} (see  Figure~\ref{fig:Bfield_hyper}).

Finally, we introduce a finite ordinary magnetic diffusivity  in the star's interior ($\eta_{\rm i} \neq 0$ in equation~\ref{eq:eta}), instead of magnetic hyper-diffusivity. These simulations correspond to Models If, Ig, and Ih of Table~\ref{tab:models}. As seen in Figure~\ref{fig:Axial_mag}, the magnetic diffusivity also makes the magnetic field evolve towards more axisymmetric configurations on the diffusive time scale, i.~e., $\mathcal{A}$ decreases substantially for larger values of $\eta_{\rm i}$, while the magnetic energy concentrates on longer wavelengths.

\subsection{Magnetic diffusivity profile in the atmosphere}\label{sec:sim_diffAtm}
\begin{figure}
    \centering
    \includegraphics[width=\columnwidth]{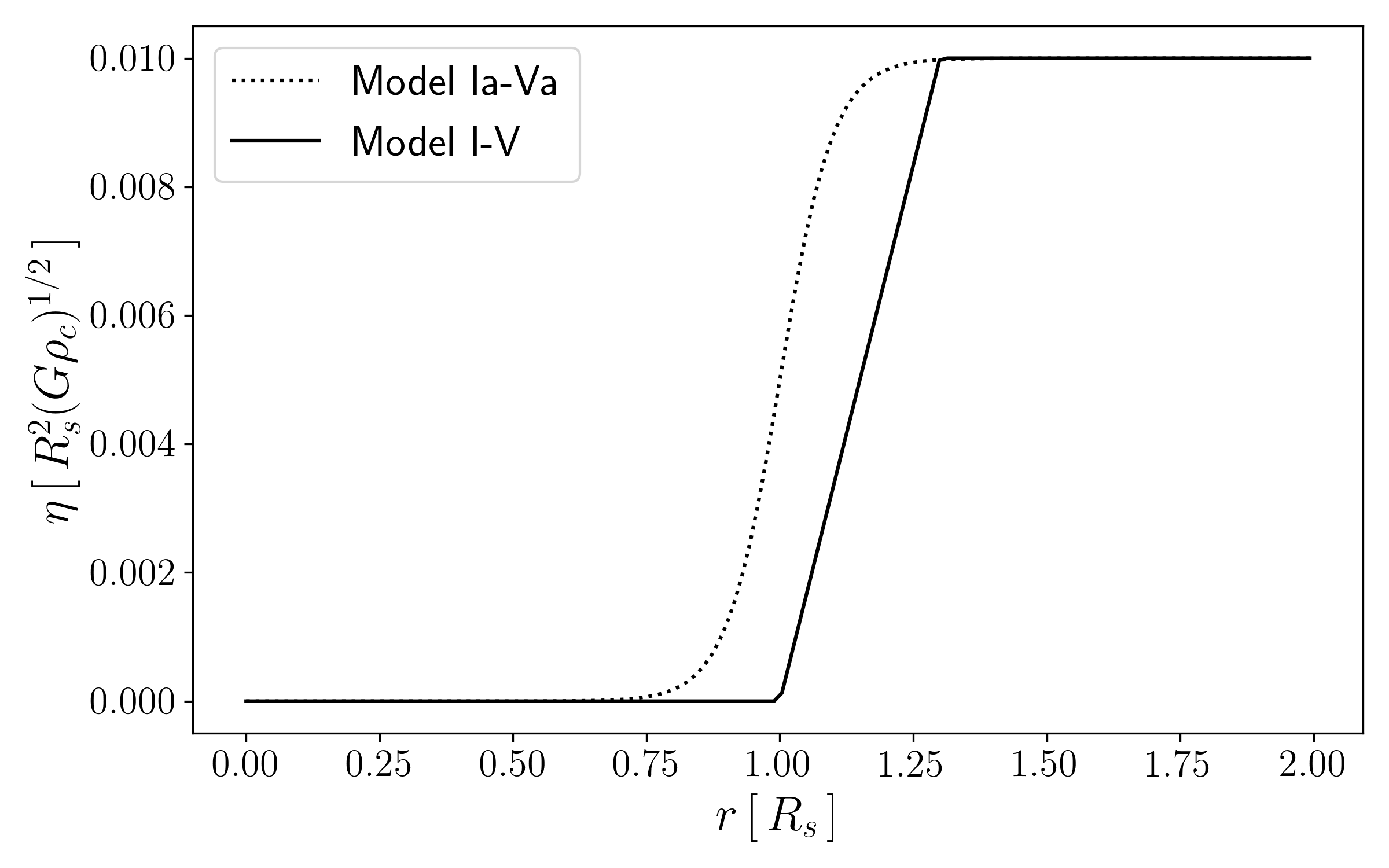}
    
    \caption{ Radial profiles of the magnetic diffusivity.} 
    \label{fig:eta}
\end{figure}

Now, we explore the effect that the transition between the vanishing magnetic diffusivity in the stellar interior and its finite value in the atmosphere has on the magnetic field evolution. For this purpose, we run more simulations in which the magnetic diffusivity changes smoothly from the stellar interior to the atmosphere (see Figure~\ref{fig:eta}):
\begin{equation}\label{eq:eta_step}
    \eta(r) = \frac{\eta_{\rm ext}}{2}\left[1+{\rm tanh} \left(\frac{r-R_s}{w}\right)\right]\, ,
\end{equation}
with $w=0.1R_s$, so it takes non-zero values in the outer layers of the star. These simulations correspond to  Models Ia (with the viscous force acting) and Va (with the hyper-diffusion scheme working) of Table~\ref{tab:models}.

\begin{figure}
    \centering

    \includegraphics[width=\columnwidth]{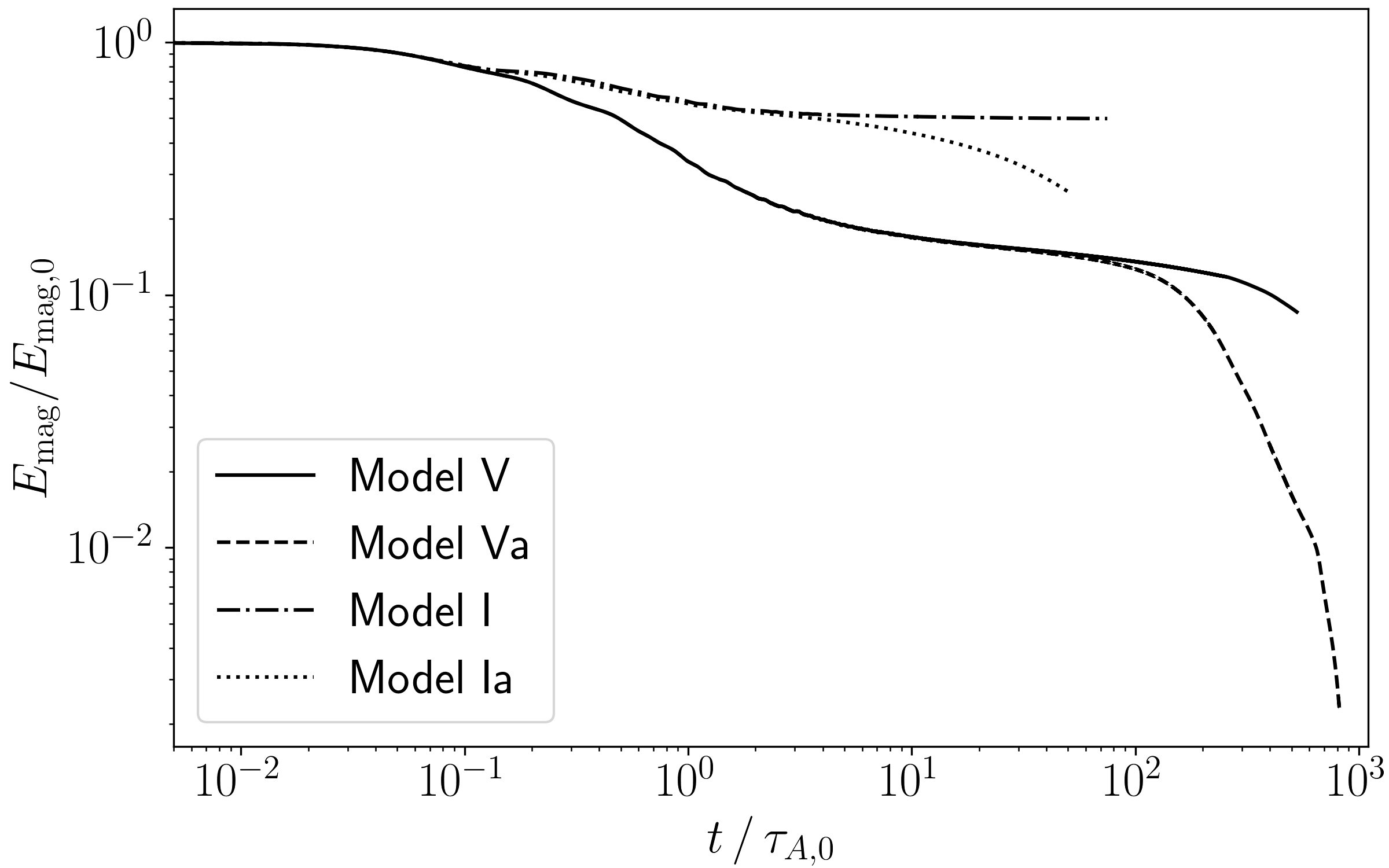}
    \caption{ Evolution of  the  total magnetic energy for the different profiles of the magnetic diffusivity plotted in Figure~\ref{fig:eta} (present only outside the star for Models I and V, and also in a surface layer in Ia and Va). Models I and Ia are done with ordinary viscosity while Models V and Va have hyper-viscosity and magnetic hyper-diffusivity.}
    \label{fig:Emag_atm}
\end{figure}
\begin{figure*}
    \centering
\subfigure[$t=192.8~\tau_{A.0}$]{ \includegraphics[width=0.3\textwidth]{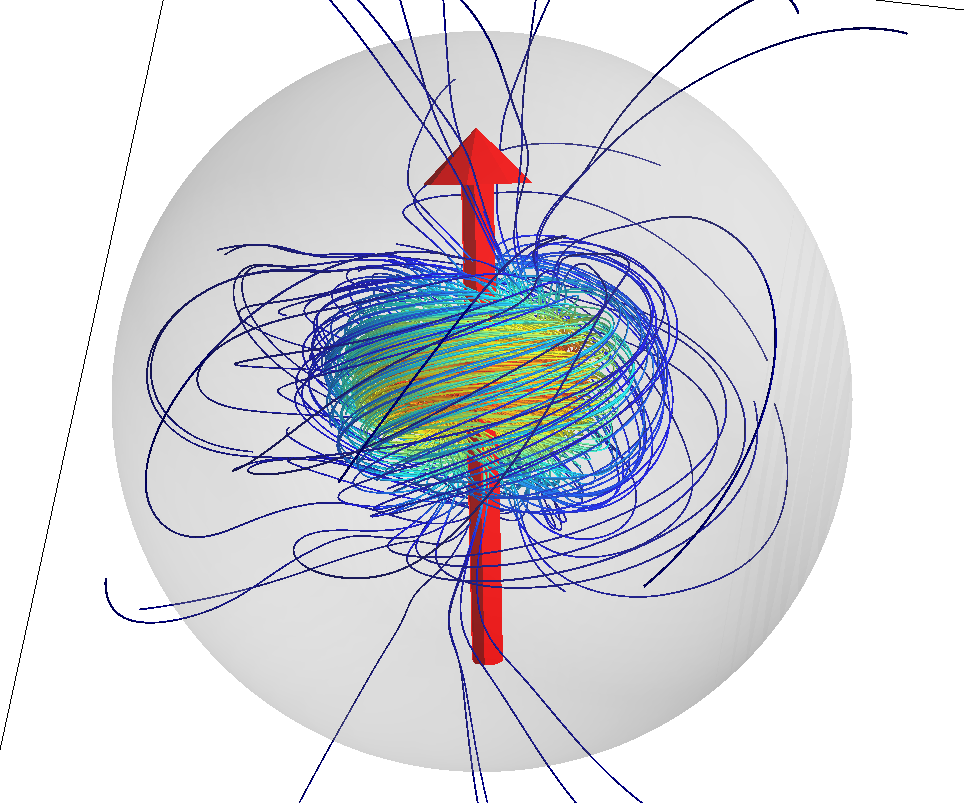}}
   \subfigure[$t=602.5~\tau_{A.0}$]{ \includegraphics[width=0.33\textwidth]{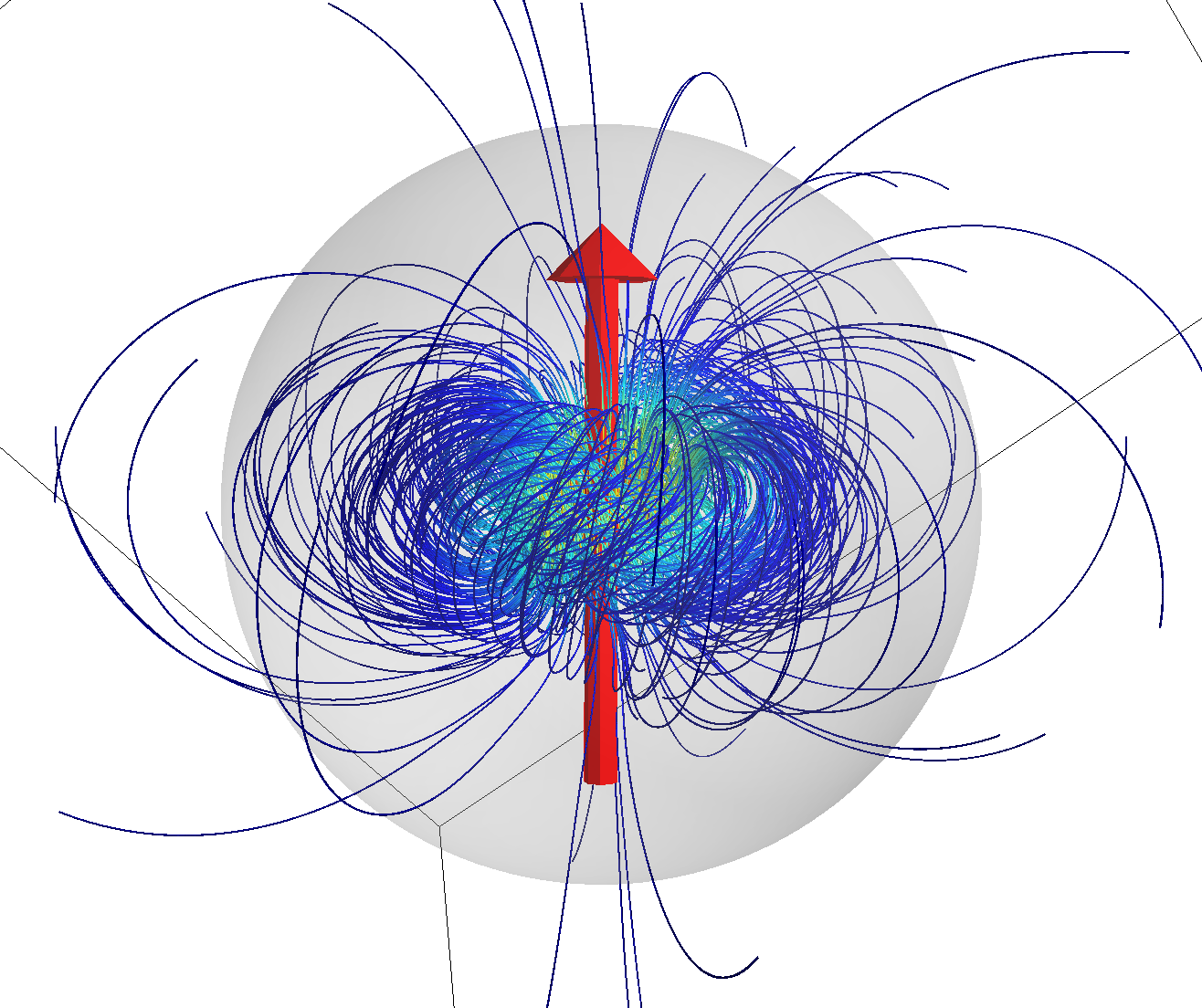}}
  \subfigure[$t=722.9~\tau_{A.0}$]{\includegraphics[width=0.33\textwidth]{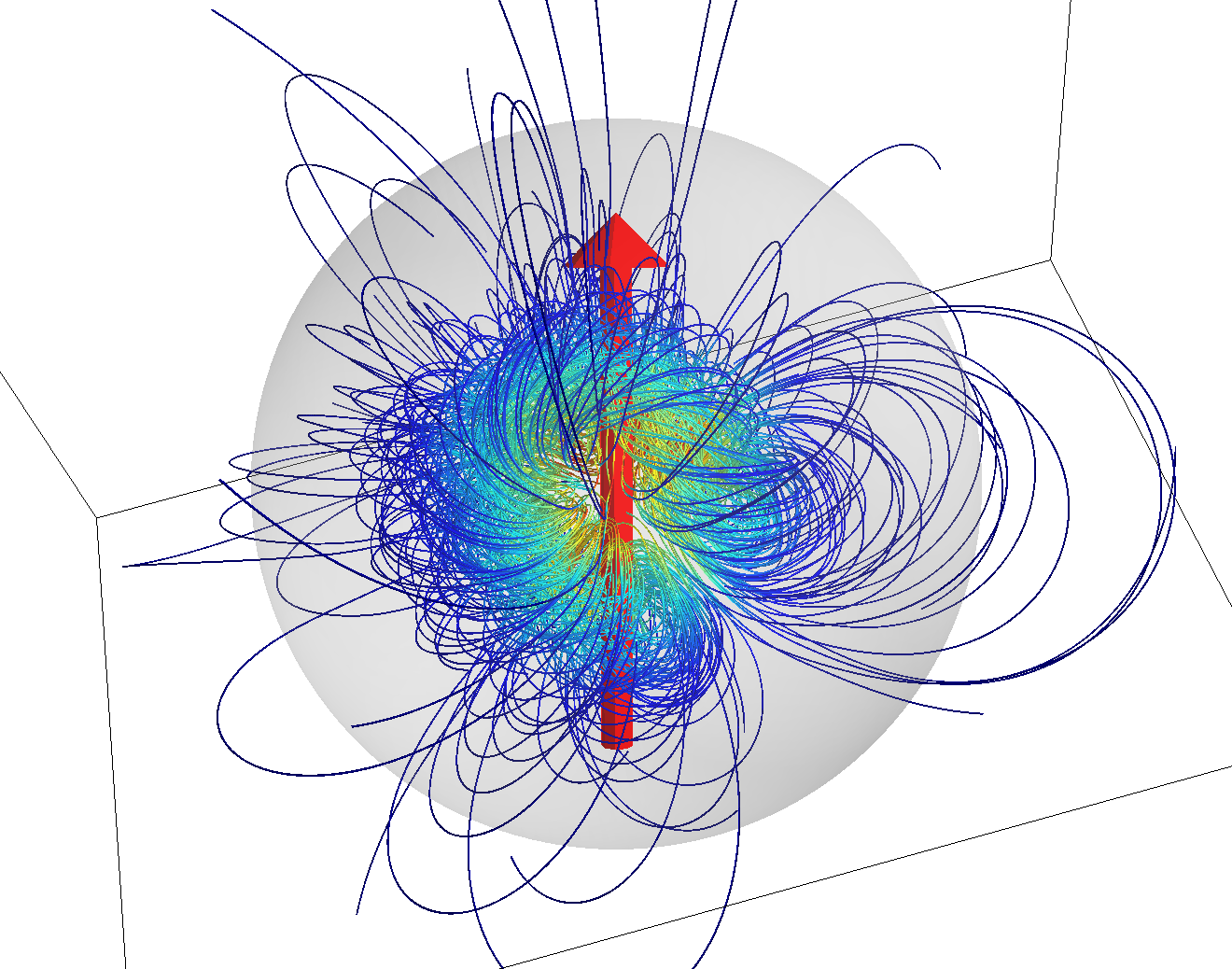}}
    \caption{3D view of the magnetic field lines for Model Va in Table~\ref{tab:models} at three different times. This simulation has $\nu_3=\eta_3=4.4\times 10^{-10} R_s^6/\tau_{A,0}$ and the magnetic diffusivity has the smooth profile of equation~(\ref{eq:eta_step}), extending into the surface layers of the star. The arrows represent the magnetic axis, $\vec{M}$, calculated at each time. The gray background in each plot corresponds to the stellar surface.}
    \label{fig:Bfield_atm}
\end{figure*}

As seen in Figure~\ref{fig:Emag_atm}, the initial settling into a hydromagnetic equilibrium state is independent of the magnetic diffusivity profile, since the magnetic field is concentrated in the stellar interior, so in this region the dissipation time scale is longer (the magnetic diffusivity is smaller). However, after a few hundred Alfv\'en times, the magnetic energy for the smoother diffusivity profile (extending farther into the star) drops substantially.

The very late-time behavior of Model Va (with the hyper-diffusion scheme turned on) is further analyzed  in Figure~\ref{fig:Bfield_atm}, showing 3D views of the magnetic field configurations at three different times. At  $t=192.8~\tau_{A,0}$, the torus around the magnetic axis inside the star is clearly visible. At $t=602.5~\tau_{A,0}$, the magnetic energy has decayed, the torus has expanded to the stellar surface, and the dipolar field configuration outside the star has become visible. Finally, at $722.9~\tau_{A,0}$, the magnetic field is deformed into a `tennis-ball'-like shape, also described in \cite{2006A&A...450.1077B}. This evolution is similar to that seen in the simulations of \cite{2004Natur...431..819B} and \cite{2006A&A...450.1077B}. The late change of the magnetic field geometry seems to be related to the fact that magnetic diffusivity is present in the outer layers of the star, not just outside.

\subsection{Dependence on the initial magnetic field configuration}\label{sim:initial_magconf}

\begin{figure}
    \centering
    \subfigure[]{\includegraphics[width=0.48\columnwidth]{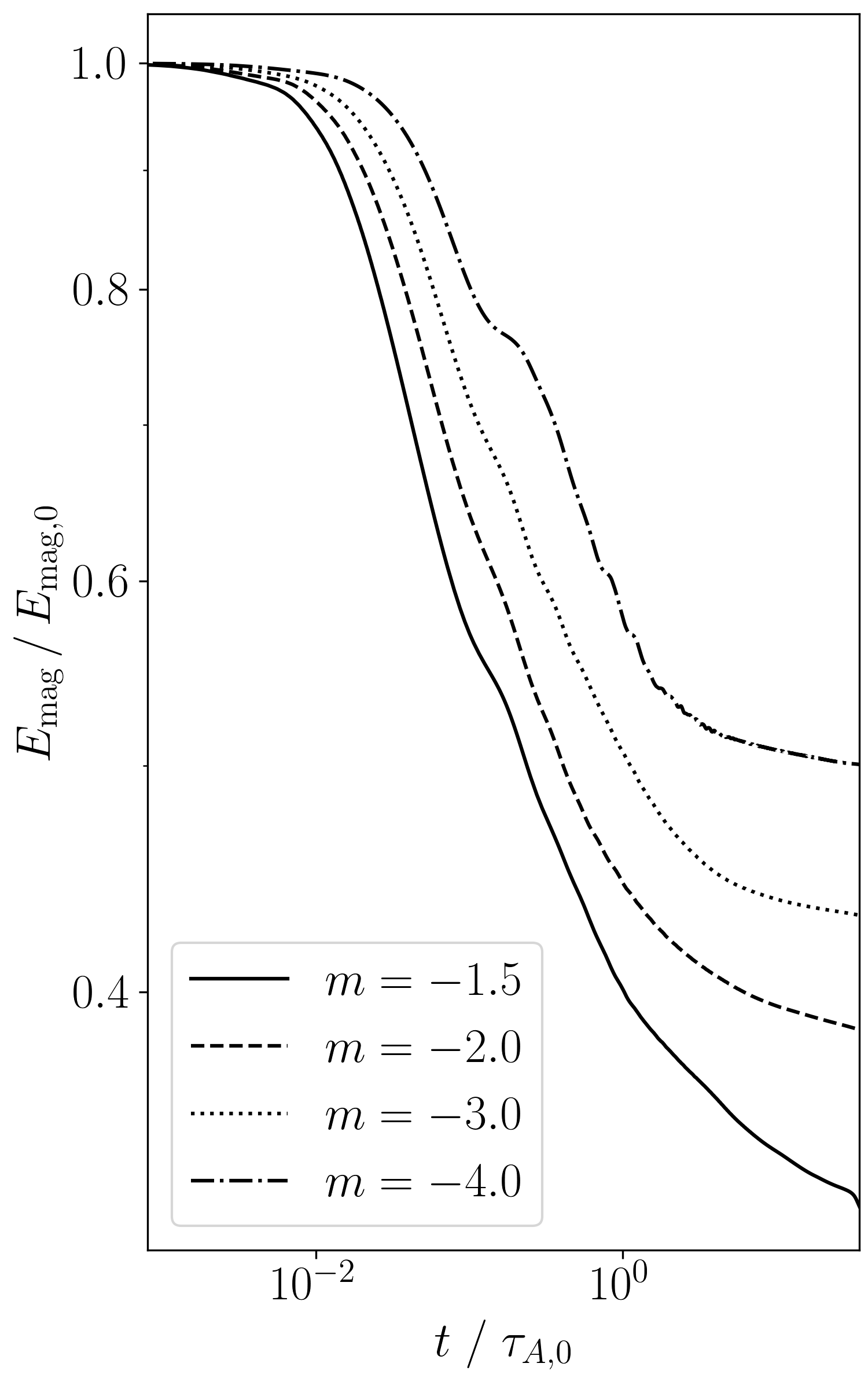}}
    \subfigure[]{\includegraphics[width=0.48\columnwidth]{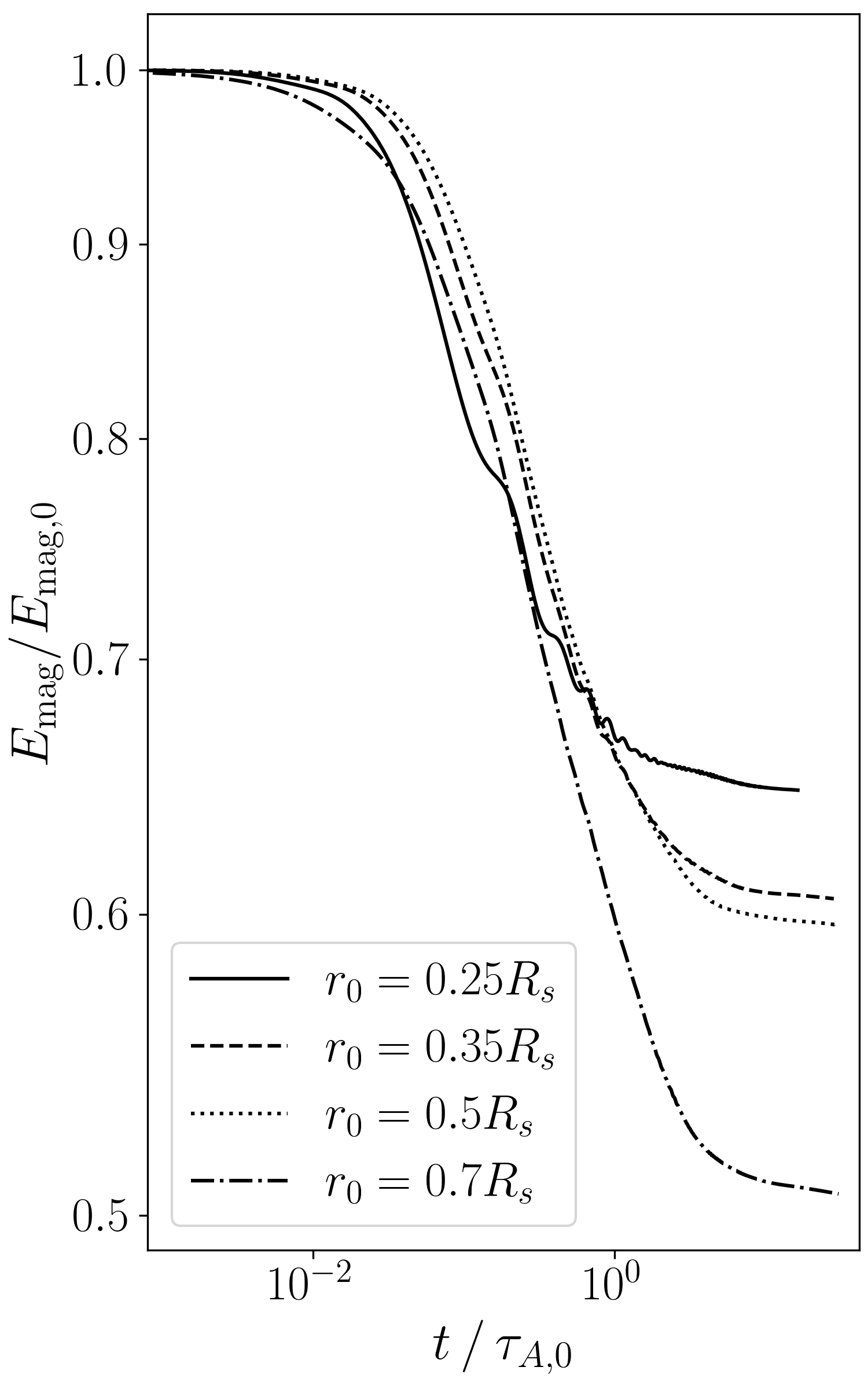}}
    
    \caption{Magnetic energy evolution for (a) models I ($m=-4$), VII ($m=-1.5$), VIII ($m=-2$) and IX ($m=-3$) and  (b) models  I ($r_0=0.25$), X ($r_0=0.35$), XI ($r_0=0.5$) and XII ($r_0=0.75$) of Table~\ref{tab:models}. These simulations have ordinary viscosity and no magnetic diffusivity (or hyper-diffusivity).
    }
    \label{fig:energy_kk}
\end{figure}

In order to test the dependence of the final outcome of the simulations on the initial magnetic field configuration,  we generate different initial random magnetic fields, first changing the value of the exponent $m$ in equation (\ref{eq:fourie_distr}) to $-1.5$, $-2$, and $-3$ (Models VII, VIII and IX of Table~\ref{tab:models}, respectively).
Smaller values of $|m|$ imply that the initial magnetic energy is more concentrated on the shortest wavelengths. On the other hand, in order to concentrate the initial magnetic field near the center of the star, we included the factor ${\rm e}^{-r^2/r_0^2}$ in the vector potential, initially taking $r_0=0.25\,R_s$, as done by \cite{2006A&A...450.1077B}. Now, we run simulations with the same random initial vector potential and the same initial total magnetic energy, but for different values of the scale in the Gaussian factor, $r_0/R_s = 0.25, 0.35, 0.5$, and $0.7$ (Models I, X, XI, and XII of Table~\ref{tab:models}, respectively). The magnetic energy evolution of these simulations, all with ordinary viscosity as the only dissipation mechanism, can be seen in Figure~\ref{fig:energy_kk}, which includes Model I for reference. In all cases, there is a rapid initial decrease of the magnetic energy, followed by a decay on a much longer time scale. This suggests that the magnetic field dynamically relaxes to an equilibrium configuration independent of the value of $|m|$ or $r_0$.

\begin{figure}
    \centering
    \includegraphics[width=0.95\columnwidth]{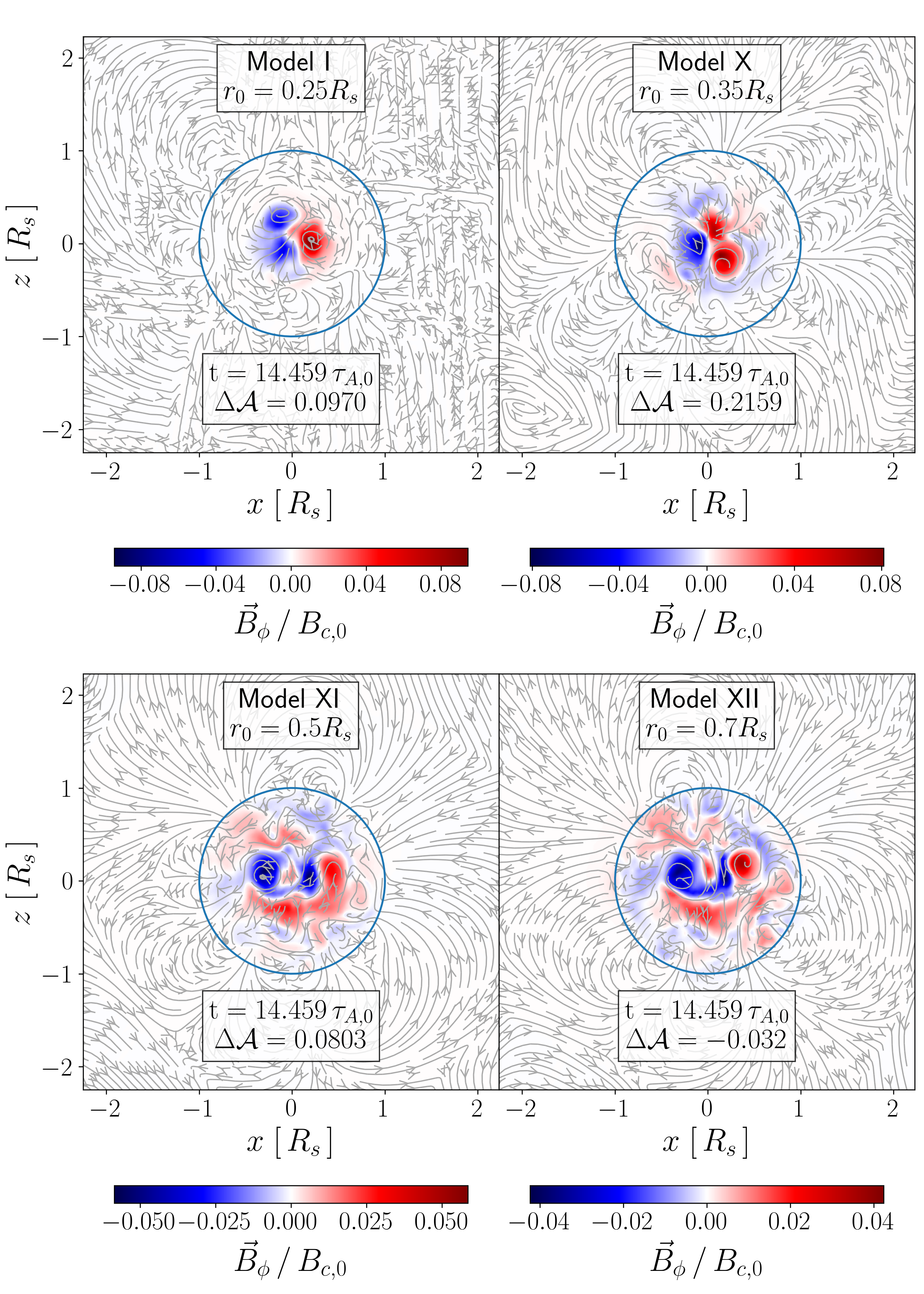}
 
    \caption{Same as Figure~\ref{fig:Bfield_snap} for different initial magnetic field concentrations $r_0$ (Models I, X, XI, and XII of Table~\ref{tab:models}) at $t\approx24.098\tau_{A,0}$. All these simulations are done with regular viscosity.}
    \label{fig:Bfield_rmag_visc}
\end{figure}

\begin{figure}
    \centering
    \includegraphics[width=0.95\columnwidth]{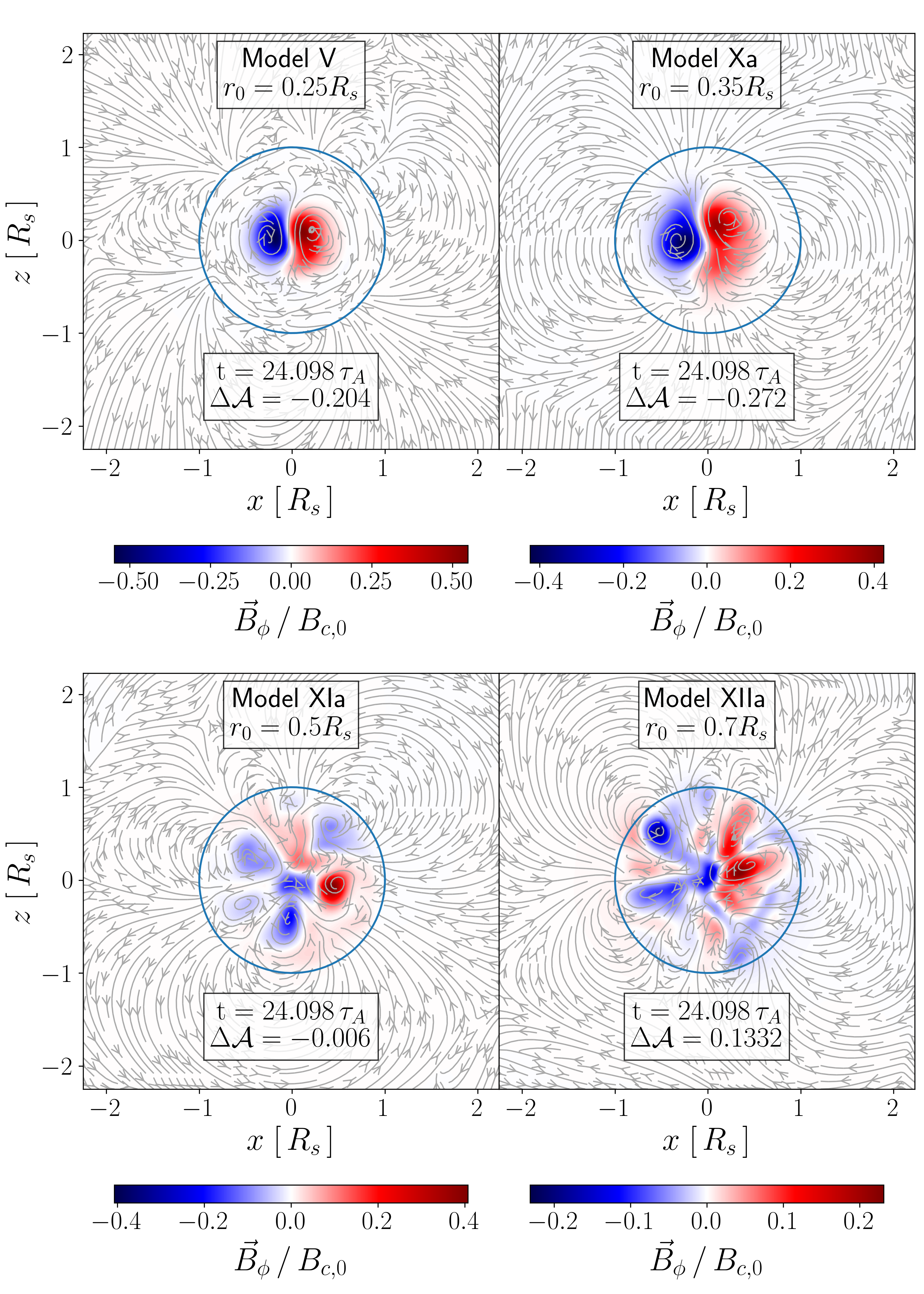}
 
    \caption{Same as Figure~\ref{fig:Bfield_snap} for different initial magnetic field concentrations $r_0$ (Models V, Xa, XIa, and XIIa of Table~\ref{tab:models}) at $t=24.098\tau_{A,0}$. All these simulations are done with $\nu_3=\eta_3=4.4\times 10^{-10} R_s^6/\tau_{A,0}$.}
    \label{fig:Bfield_rmag_hyper}
\end{figure}

For all the initial conditions tried here, and always with ordinary viscosity as the only dissipation mechanism, the magnetic field remains non-axisymmetric (see Figure~\ref{fig:Bfield_rmag_visc}). 
However, \cite{2006A&A...450.1077B}, using a hyper-diffusion scheme, found a critical value $r_0/R_s\sim 0.57$ below which the initially random field evolved to a stable axially symmetric configuration. In order to reproduce these results, we repeat the simulations of models X ($r_0=0.35R_s$), XI ($r_0=0.5R_s$), and XII ($r_0=0.75R_s$), but turning on the hyper-diffusion scheme (Models Xa, XIa, and XIIa of Table~\ref{tab:models}). Figure~\ref{fig:Bfield_rmag_hyper} shows snapshots of the magnetic field at $t=24\tau_{A,0}$ for these models and model V. The change of the asymmetry parameter, $\Delta \mathcal{A}$, is shown at the bottom of each panel. For $r_0/R_s=0.25$ and $0.35$, the final magnetic field equilibrium is formed by a  torus-like structure, and it has evolved to a more axisymmetric state than the initial one (i.~e., the change of the asymmetry parameter is negative). On the other hand, for  $r_0/R_s=0.5$ and $0.75$, the magnetic field configuration is much less ordered, and the asymmetry parameter remains nearly the same or increases. Thus, for our simulations with hyper-diffusion, the critical value of $r_0/R_s$ for the formation of an axisymmetric configuration lies between $0.35$ and $0.5$, not too different from the results of \cite{2006A&A...450.1077B}.

%

\subsection{Dependence on the star's stratification: simulations in barotropic stars}\label{sec:sims_BarStar}

Up to this point, we have discussed simulations in stably stratified stars ($\Gamma=5/3>\gamma=4/3$) with random initial magnetic field configurations. Independent of the initial conditions, their magnetic field evolved to an equilibrium configuration. 
Various papers have suggested that the stable stratification of the star is an important ingredient for the stability of the magnetic field \citep{2009MNRAS.397..763B,2009A&A...499..557R,2012MNRAS.424..482L}, and \cite{2015MNRAS.447.1213M} supported this hypothesis through a systematic study of the evolution of the magnetic field in barotropic stars with both disordered (random) and ordered (axisymmetric) initial magnetic field configurations, without finding any stable equilibria. 

\begin{figure}
    \centering
    
    \subfigure[]{\includegraphics[width=0.48\columnwidth]{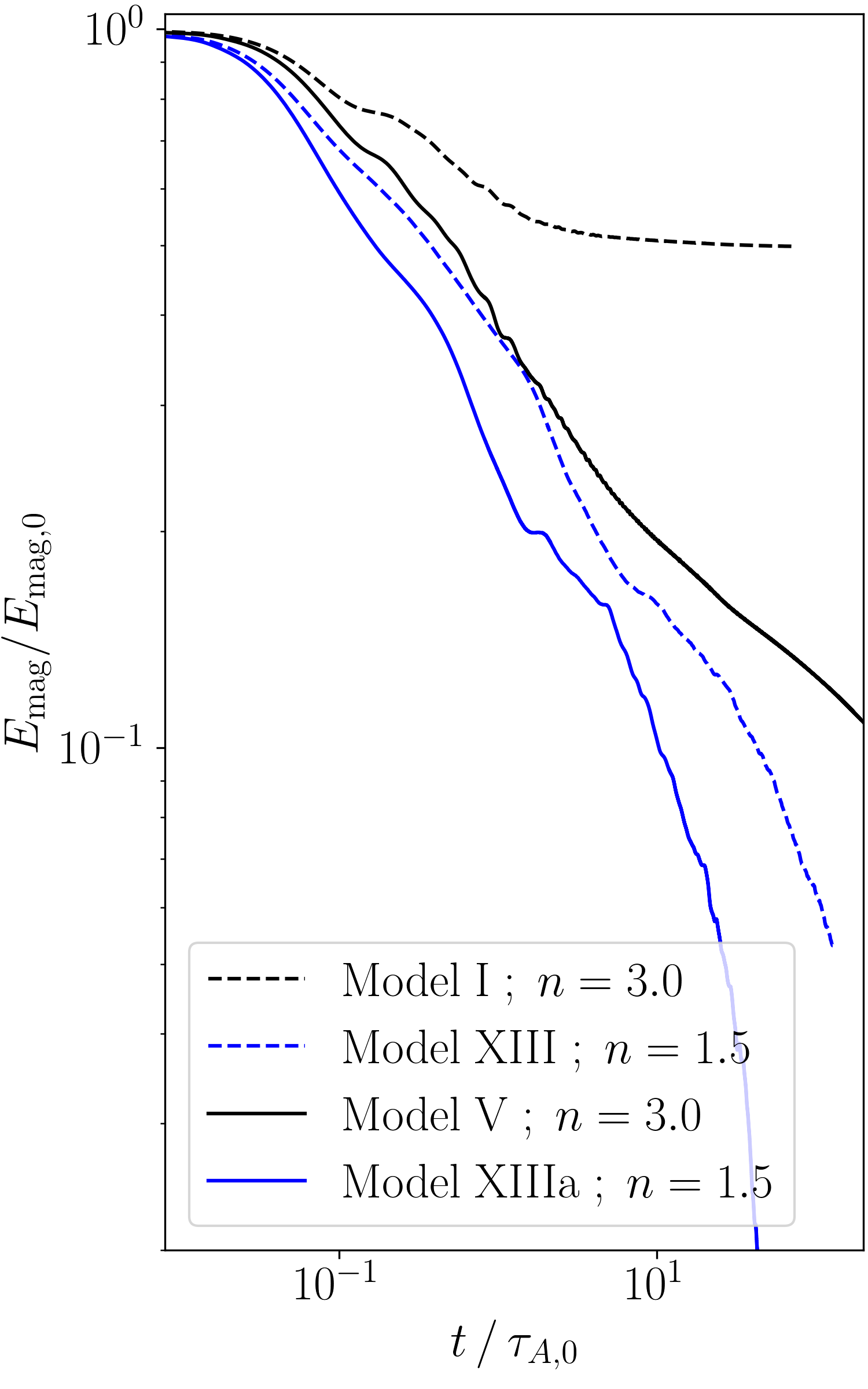}}
    \subfigure[]{\includegraphics[width=0.48\columnwidth]{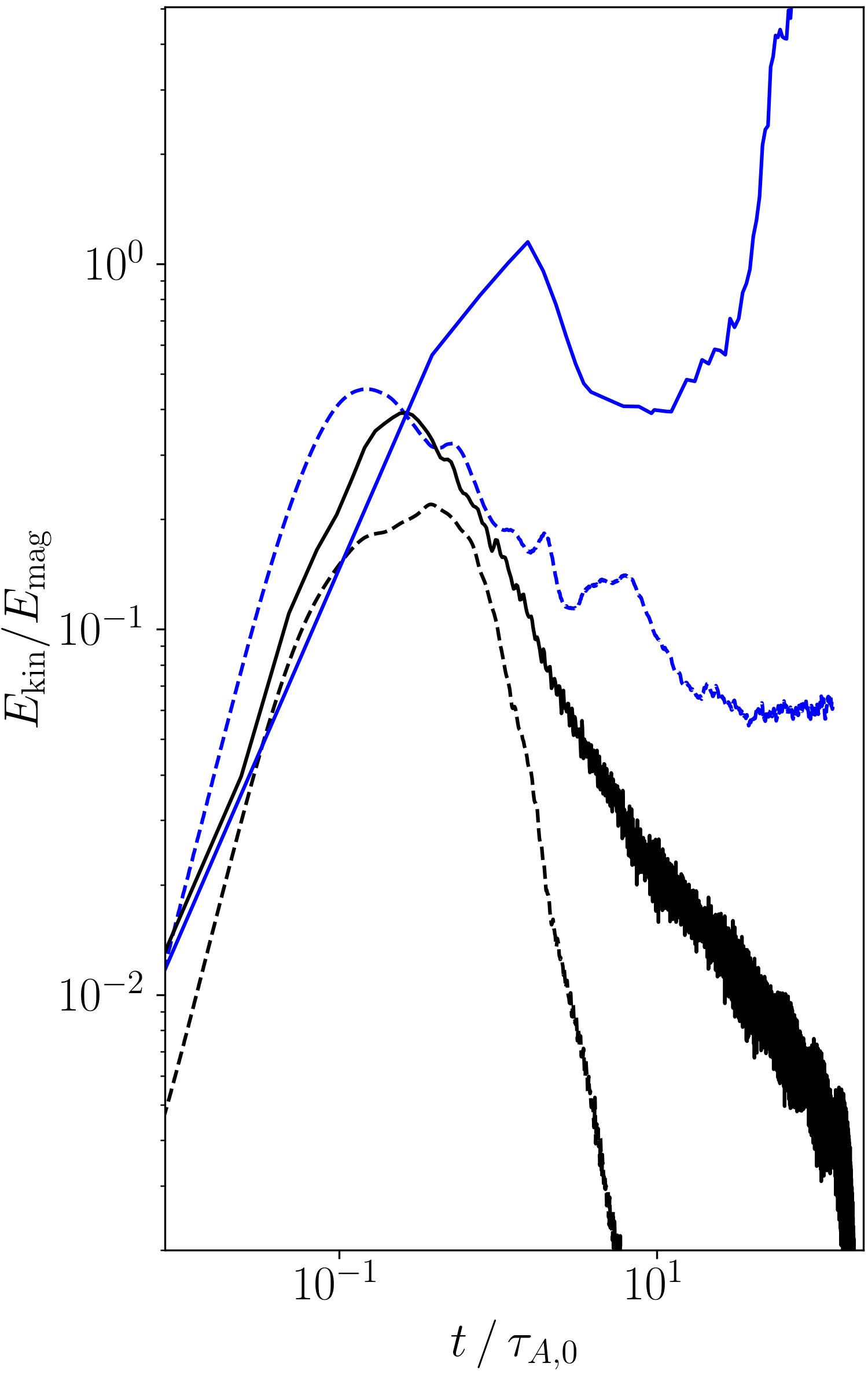}}
   
    \caption{Time evolution of the (a) magnetic energy  and (b) the ratio between  the total kinetic energy and magnetic energy for stably stratified ($\gamma=4/3$, black lines) and barotropic ($\gamma=5/3$, blue lines) stars.
    The parameters used in these simulations correspond to the ones of Model I/V  and Model XIII/XIIIa  of Table~\ref{tab:models} for the stratified ($n=3.0$) and  barotropic stars ($n=1.5$), respectively. Models I and XIII (dashed lines) are done with ordinary viscosity, while Models V and XIIIa (solid lines) are done with hyper-viscosity and magnetic hyper-diffusivity. }
    \label{fig:Emag_bar}
\end{figure}

To validate the application of the {\sc Pencil code} in this scenario, we run similar simulations, changing the initial polytropic index to $n=1.5$ ($\Gamma=\gamma=5/3$), so the entropy is uniform inside the star (see Figure~\ref{fig:init_cond}). Figure~\ref{fig:Emag_bar} shows the evolution of the magnetic energy and the ratio of  kinetic and magnetic energy for two different random initial magnetic field configurations obtained with the procedure described in section~\ref{sec:randomField}, one run with just ordinary viscosity and the other introducing the hyper-diffusion scheme (Models XIII and XIIIa of Table~\ref{tab:models}, respectively). For comparison,  we show the evolution of the same initial magnetic fields in a stably stratified star. For the barotropic simulations, the magnetic energy decreases much more quickly than in the stably stratified cases, while the ratio between the kinetic and magnetic energy does not become a decreasing function of time, indicating that the barotropic star does not approach a hydromagnetic equilibrium.

\begin{figure*}
    \centering
 \includegraphics[width=0.98\textwidth]{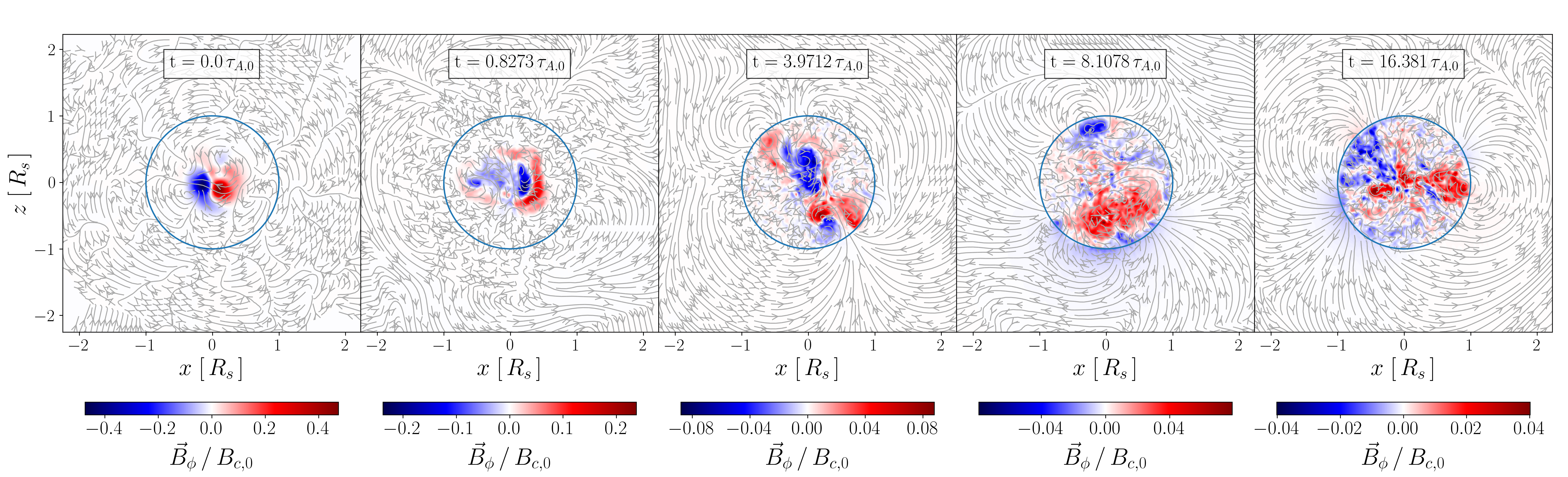}
     
    \caption{Snapshots of the magnetic field for a barotropic star (Model XIII, with ordinary viscosity, of Table~\ref{tab:models}) at times as indicated. The color scale and field lines are the same as in Figure~\ref{fig:Bfield_snap}.} 
    \label{fig:Bfield_barotropic}
\end{figure*}

Figure~\ref{fig:Bfield_barotropic} shows the evolution of an initially random magnetic field in the barotropic star, with the vertical axis at each time aligned with the magnetic axis $\vec{M}$.  It can be seen that the magnetic field does not reach an ordered equilibrium state, but expands to the surface of the star, probably because of magnetic buoyancy, which in this case is not counteracted by an entropy gradient \citep{2009A&A...499..557R}.  
Thus, this and many other numerical experiments conducted by us (not shown here) confirm  the results of \cite{2015MNRAS.447.1213M} (and \citealt{2012MNRAS.422..619B}) in the sense that barotropic stars do not appear to be able to support any stable magnetic equilibria.

\section{Conclusions}\label{sec:conclusion}

We have modeled the evolution of random initial magnetic fields in the interiors of both stably stratified and barotropic stars, solving the MHD equations with the {\sc Pencil Code}. 

We confirmed the main results of \cite{2004Natur...431..819B} and \cite{2006A&A...450.1077B}, in the sense that a random initial magnetic field in a stably stratified star with adiabatic index $\Gamma=5/3$ and polytropic index $n=3$ ($\gamma=4/3<\Gamma$) relaxes to a stable equilibrium configuration. 
On the other hand, no stable magnetic field configurations were reached in barotropic stars ($n=1.5$ and $\gamma=5/3=\Gamma$), confirming that a star's stratification is a crucial ingredient for the stability of its magnetic field \citep{2009A&A...499..557R,2015MNRAS.447.1213M}.  

Through the calculation of the asymmetry parameter defined in equation (\ref{eq:axial_param}), we found that the evolution towards a roughly axisymmetric magnetic field equilibrium, reported by \cite{2004Natur...431..819B} and \cite{2006A&A...450.1077B}, occurs only if magnetic diffusivity or magnetic hyper-diffusivity is present in the simulations, not if viscosity or hyper-viscosity is the only dissipation mechanism. We believe that regular viscosity more accurately mimics the real physics inside stars, in which the equilibrium is likely achieved through phase mixing, which, like viscosity, stops acting once the motions are damped. If this is the case, the final (quasistationary) magnetic field configuration is likely to be about as asymmetric as the initial magnetic field. We note, however, that the values used for the viscosity in our simulations are much larger than in real stars, so we can only speculate that the final results will hold also for more realistic values.

Once the magnetic field has relaxed to a stable equilibrium, it evolves diffusively. Ohmic diffusion and thermal diffusion appear to be too slow to significantly affect the evolution of the magnetic field in the lifetime of Ap stars \citep{2009A&A...499..557R}, although there are observational indications of moderate magnetic field decay beyond  flux conservation during stellar expansion, for yet unidentified reasons \citep{2007A&A...470..685L,2008A&A...481..465L, 2016A&A...592A..84F}. 

We note that we have assumed a non-rotating star, thus, the results of this work apply when the Alfv\'en frequency is larger than the star's rotation rate, namely, for slow rotators or high magnetic field strengths. The final magnetic field configuration could be affected by the star's rotation if the star's rotation rate is larger than the Alfv\'en frequency.

%

\section*{Acknowledgements}

The authors are grateful to J. Braithwaite and H. Spruit for useful discussions. This work was supported by FONDECYT projects 3190172 (L.B.), 1201582 (A.R.), and 1190703 (J.A.V.).
 J.A.V. thanks for the support of CEDENNA under CONICYT grant AFB180001. M.E.G. acknowledges partial support from Russian Foundation for Basic Research [Grant No.\ 19-52-12013].
Some of the calculations presented in this work were performed on the Geryon computer at the Center for Astro-Engineering UC. BASAL CATA PFB-06, the Anillo ACT-86, FONDEQUIP AIC-57, and QUIMAL 130008 provided funding for several improvements to the Geryon cluster.

\section*{Data Availability}

The data underlying this article will be shared on reasonable request to the corresponding author.


\bibliographystyle{mnras}
\bibliography{biblio}



\appendix

\section{Set-up of the stellar model}\label{ap:star}  
In order to determine the gravitational potential, as well as the initial density and pressure profiles, we solve the (non-magnetic) hydrostatic equilibrium equations,
\begin{eqnarray}
    \frac{d p}{d r} &=& -\rho\frac{d \Phi }{d r}\label{eq:HydEqs_a},\\
    \frac{1}{r^2}\frac{d }{dr}\left(r^2 \frac{d \Phi }{d r}\right) &=& 4\pi G \rho \label{eq:HydEqs_b}, 
\end{eqnarray}
where $p(r)$ and $\rho(r)$ are the pressure and density of an unmagnetized, spherically symmetric star, and $r$ is the radial coordinate, together with the pressure-density relation
\begin{equation}
    P(r)= \left\{ 
     \begin{aligned}
     K\rho(r)^\gamma \qquad &\mathrm{for} & \quad
r \leq R_s \\
   (\mathcal{R}/\mu) T_{\rm atm}(r) \rho(r)  \qquad &\mathrm{for} & \quad
 R_s < r < R_s+\Delta_r \\
 (\mathcal{R}/\mu) T_{\rm out} \rho(r)  \qquad &\mathrm{for} & \quad
  r \geq R_s+\Delta_r \, ,
     \end{aligned} 
     \right.
\end{equation}
with  $T_{\rm atm}(r)= T_s+(T_{\rm out}-T_s)\left(r-R_s \right)/\Delta_r$, where $T_s$ is the star's surface temperature, $T_{out}$ 
is the temperature in the outer atmosphere, $R_s$ is the stellar radius, and  $\Delta_r= 0.3~R_s$. The equation of state over the whole simulation box is the ideal gas equation of state (equation~\ref{eq:pres_def}). Then, the temperature inside the star ($r\leq R_s$) can be obtained from equation~(\ref{eq:pres_def}) and the entropy from equation~(\ref{eq:tem_def}).

Given certain values for the stellar mass, $M_s$, and radius, $R_s$,  the value of the constant $K = c_{\rm s,c}^2\rho_c^{1-\gamma}/\Gamma$ is set. For a stably stratified star with $\gamma=4/3$, the following scalings  are satisfied for the central density, $\rho_c$, central sound speed, $c_{s,c}$, central temperature, $T_c$ and total gravitational energy, $E_{\rm grav}$: 
$$ \frac{\rho_c}{\rm g/cm^3} = 19.60\frac{M_s}{M_\odot}\frac{R_\odot^3}{R_s^3},\qquad  \qquad \frac{c_{s,c}}{10^7 {\rm cm/s}}= 4.19 \sqrt{\frac{M_s}{M_\odot}\frac{R_\odot}{R_s}},$$
$$\frac{T_c}{10^6\,\mathrm{K}}= 7.61 \frac{M_s}{M_\odot}\frac{R_\odot}{R_s},\qquad  \qquad \frac{|E_{\rm grav}|}{10^{48}\,\mathrm{ergs}} = 1.79 \left(\frac{M_s}{M_\odot}\right)^2\frac{R_\odot}{R_s}\, , $$
while, for a barotropic star with $\gamma=5/3$, these are:
$$ \frac{\rho_c}{\rm g/cm^3} = 7.53\frac{M_s}{M_\odot}\frac{R_\odot^3}{R_s^3},\qquad  \qquad \frac{c_{s,c}}{10^7 {\rm cm/s}}= 4.06 \sqrt{\frac{M_s}{M_\odot}\frac{R_\odot}{R_s}},$$
$$\frac{T_c}{10^6\,\mathrm{K}}= 7.14 \frac{M_s}{M_\odot}\frac{R_\odot}{R_s},\qquad  \qquad \frac{|E_{\rm grav}|}{10^{48}\,\mathrm{ergs}} = 1.24 \left(\frac{M_s}{M_\odot}\right)^2\frac{R_\odot}{R_s}. $$

\section{Test simulations  }\label{ap:test_sims}
%
\begin{table}
    \centering
    \addtolength{\tabcolsep}{-1.6pt}  
    \begin{tabular}{cccccc}
    \hline
      Model   & $L_{\rm box}$ & Resolution & $\nu$ &  $\nu_3$ & $\eta_3$ \\
       
        &  $[R_s]$   &    &  $[R_s^2/\tau_{\rm A,0}]$& $[R_s^6/\tau_{\rm A,0}]$ & $[R_s^6/\tau_{\rm A,0}]$  \\ \hline
      
      AIa & $4.5$  &  $ 128^3$  &$ 0.0 $ & $2.2\times 10^{-10}$ & 0.0 \\
      
      AIb & $4.5$  &  $ 128^3$  &$ 0.0 $ & $4.4\times 10^{-10}$ & 0.0 \\
      
      AIc & $4.5$  &  $ 128^3$  &$ 0.0 $ & $2.2\times 10^{-9}$ & 0.0 \\
      
      AId & $4.5$  &  $ 128^3$  &$ 0.0 $ & $4.4\times 10^{-9}$ & 0.0 \\
      
      AII & $3.1$  &  $ 88^3$  &$ 0.0 $ & $4.4\times 10^{-10}$ & $4.4\times 10^{-10}$ \\
         
      AIII & $5.9$  &  $ 168^3$  &$ 0.0 $ & $4.4\times 10^{-10}$ & $4.4\times 10^{-10}$ \\
        
      AIV & $4.5$  &  $ 96^3$ & $ 0.0 $ & $4.4\times 10^{-10}$ & $4.4\times 10^{-10}$ \\
        
      AV & $4.5$  &  $ 128^3$ & $ 0.0 $ & $4.4\times 10^{-10}$ & $4.4\times 10^{-10}$ \\
      
      AVI & $4.5$  &  $ 256^3$ &$ 0.0 $ & $4.4\times 10^{-10}$ & $4.4\times 10^{-10}$ \\
    \hline
        
    \end{tabular}
    \caption{Parameters for the  test simulations. }
   \label{tab:models_appendix}
\end{table}
\begin{figure}
    \centering
    \includegraphics[width=0.95\columnwidth]{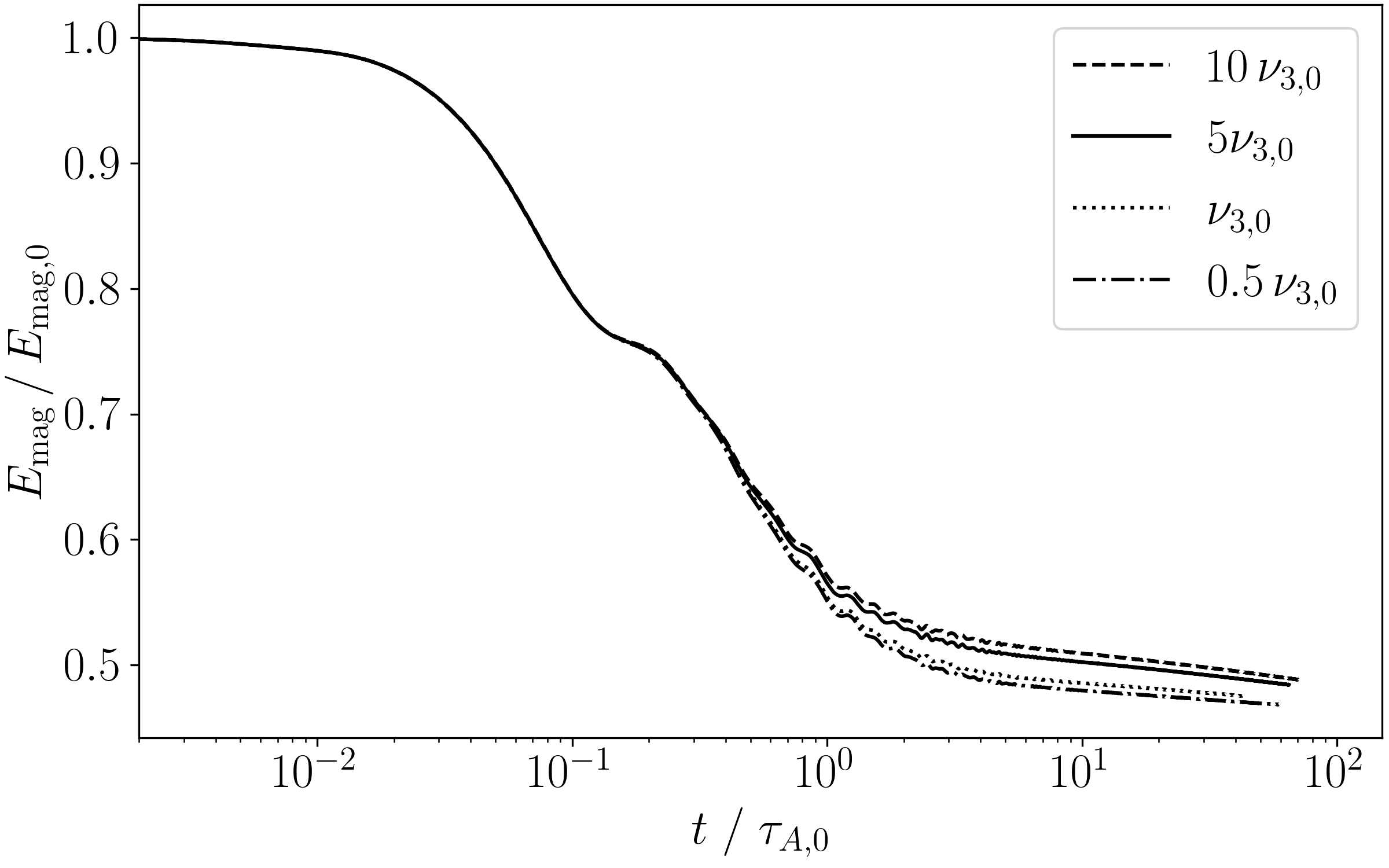}
    \caption{  Magnetic Energy evolution for four different values for the hyper-viscosity coefficient (Models AIa, AIb, AIc and AIb of Table~\ref{tab:models_appendix} with $\nu_{3,0}=4.4\times 10^{-10} R_s^6/\tau_{A,0}$).
    }
    \label{fig:Emag_hyper-visc}
\end{figure}
\begin{figure}
    \centering
    \includegraphics[width=0.95\columnwidth]{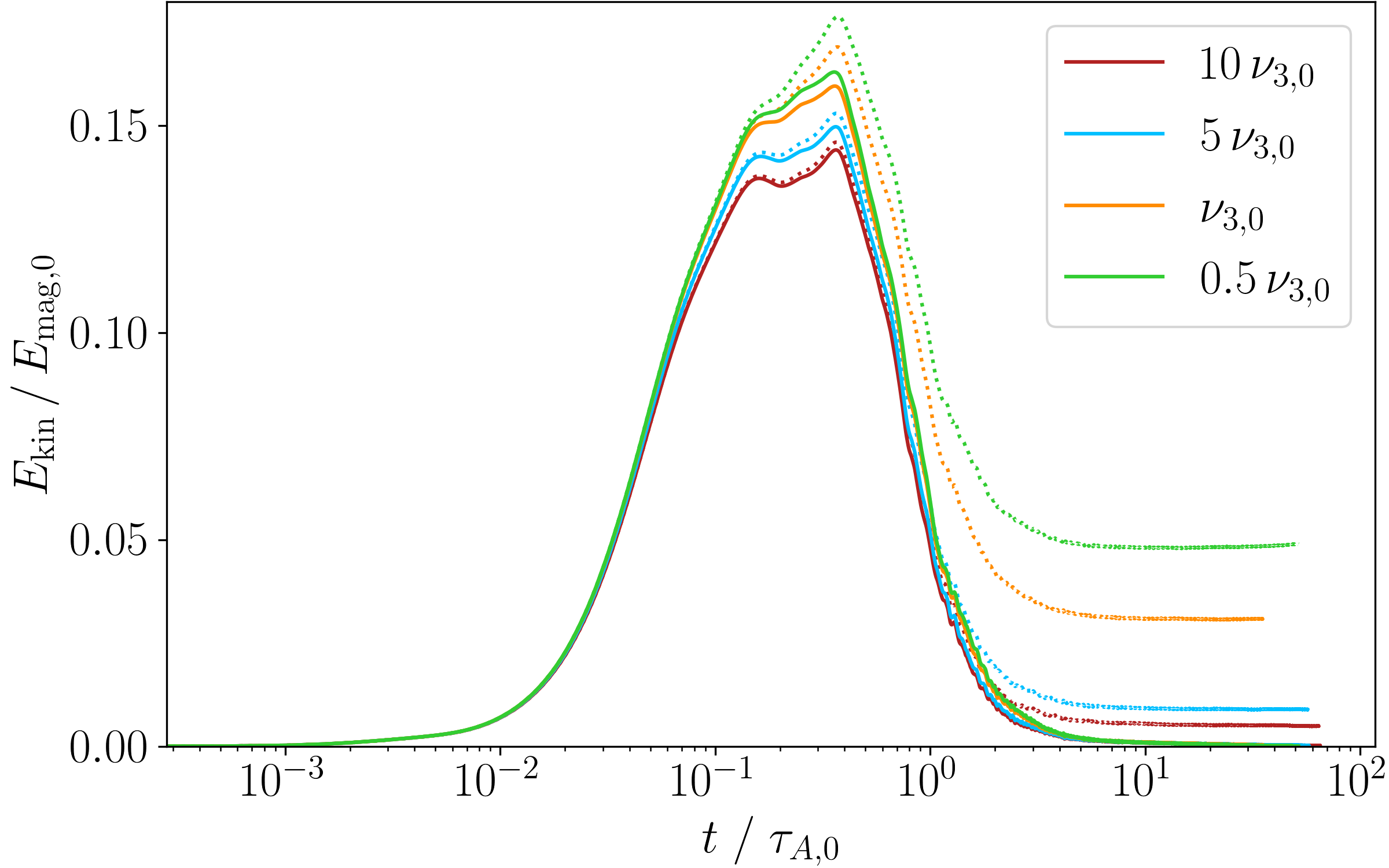}
    \caption{Same as Figure~\ref{fig:Ekin_visc} but  for simulations done with just hyper-viscosity and no magnetic diffusivity inside the star (Models AIa, AIb, AIc, and AId from Table~\ref{tab:models_appendix} and with  $\nu_{3,0}=4.4\times 10^{-10} R_s^6/\tau_{A,0}$).
    }
    \label{fig:Ekin_hyper-visc}
\end{figure}

In this appendix, we test the accuracy of the code for the simulations presented in this paper.

We start performing a similar analysis to the one done in section~\ref{sec:sim_visc}, but for simulations run with just hyper-viscosity as the dissipative mechanism for the kinetic energy. We want to determine the value for the hyper-viscosity coefficient, $\nu_3$, from which the numerical dissipation is negligible. As seen in Figure~\ref{fig:Emag_hyper-visc}, the magnetic energy evolves almost independent of the value of $\nu_{3}$. But, from the comparison of the kinetic energy calculated with equations (\ref{eq:ekin}) and numerical integrating equation (\ref{eq:ekin}) (see Figure~\ref{fig:Ekin_hyper-visc}), the most reliable simulations are the ones with $\nu_3>2.2\times 10^{-9} R_s^6/\tau_{A,0}$.

Next, we  run test simulations with a smaller ($L_{\rm box}= 3.1\,R_s$) and a larger simulation box ($L_{\rm box}= 5.9~R_s$), maintaining the same grid spacing ($\Delta x = 0.035~R_s$) and initial set-up of Model V of Table~\ref{tab:models} (Models AII and AIII of Table~\ref{tab:models_appendix}, respectively). The evolution of the magnetic field is found to be insensitive to the size of the box, so we conclude that the use of periodic boundary conditions is not affecting the evolution of the magnetic field inside the star. 

\begin{figure}
    \centering
    \subfigure[]{ \includegraphics[width=0.485\columnwidth]{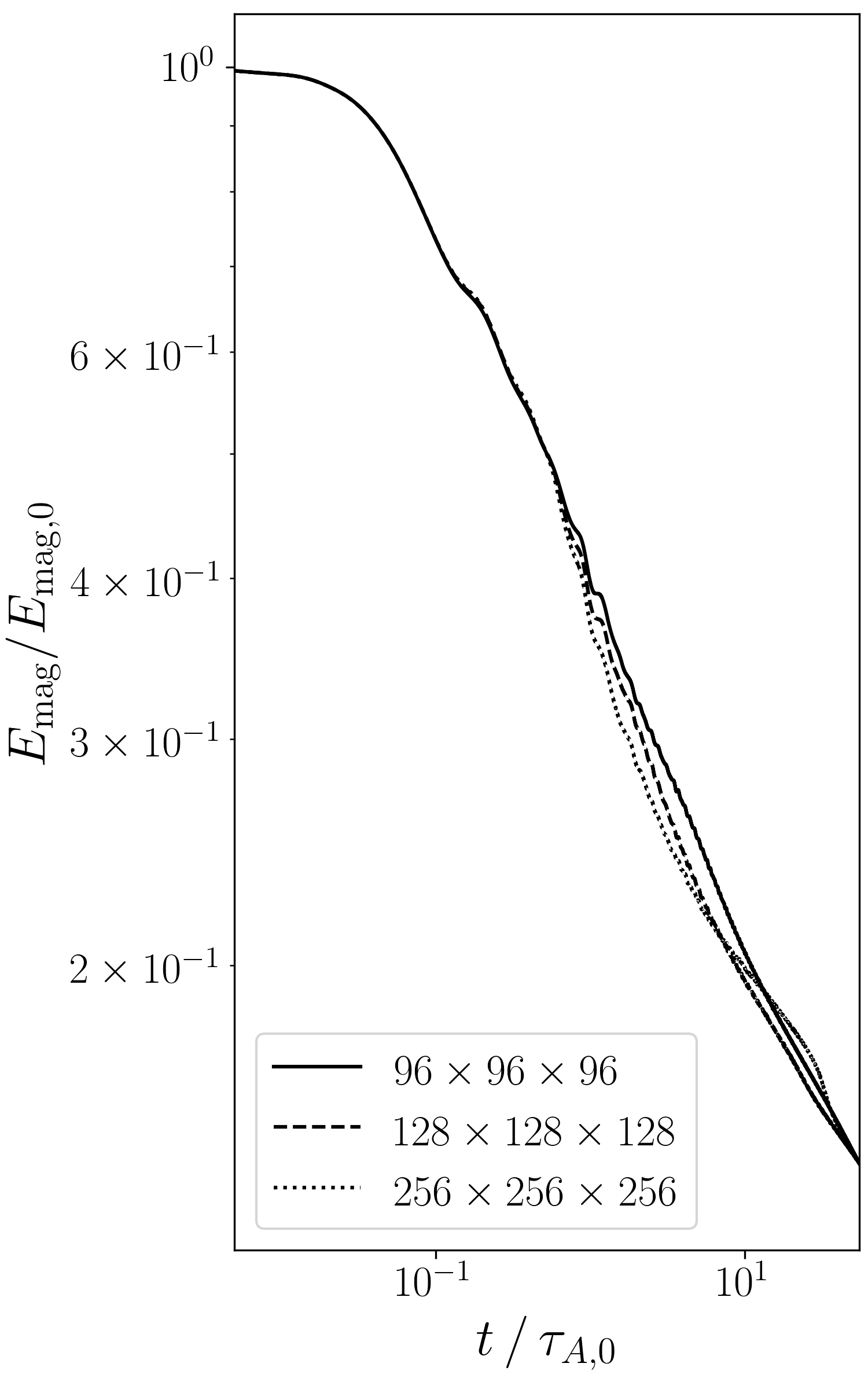}}
    \subfigure[]{ \includegraphics[width=0.48\columnwidth]{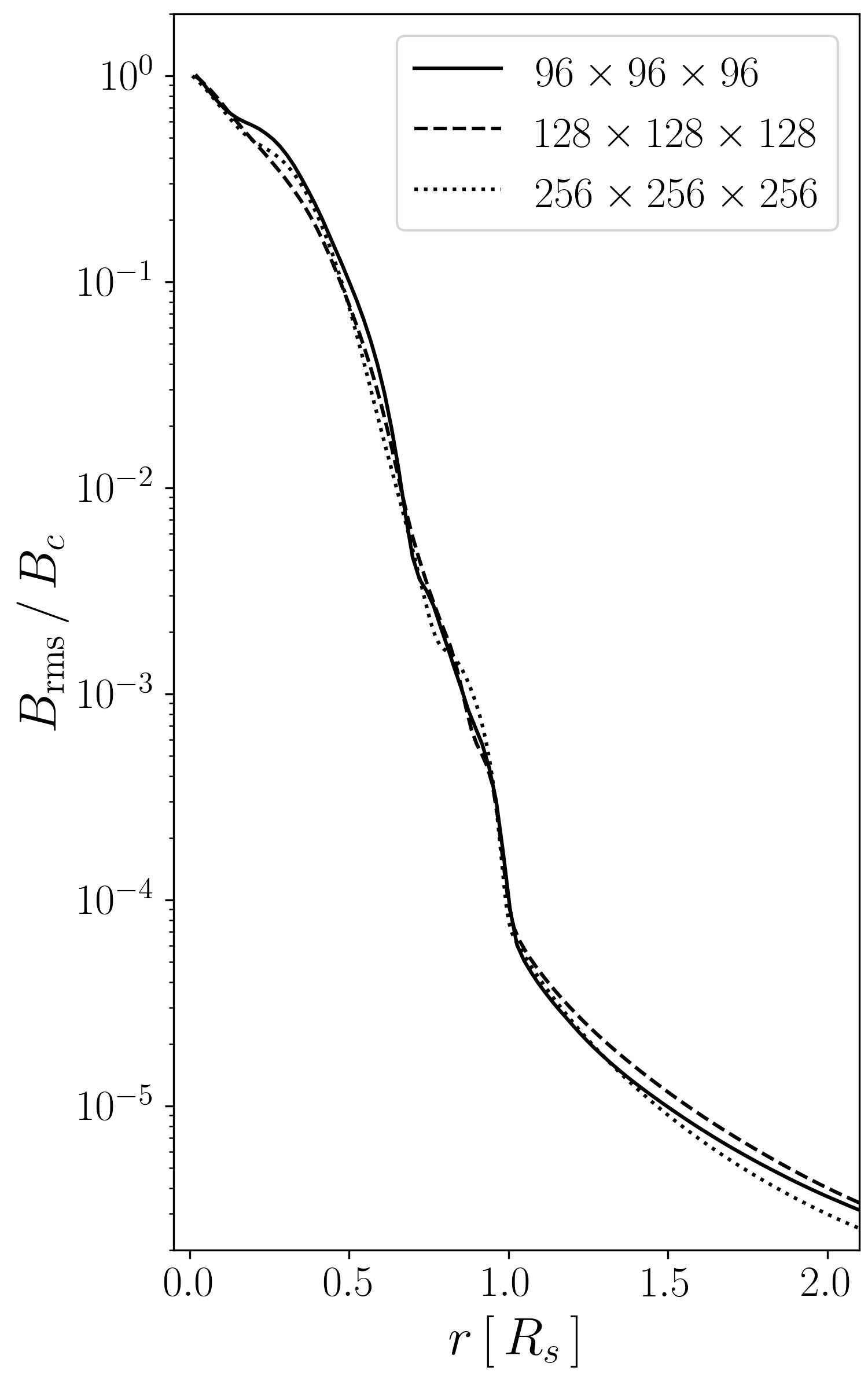}}
    
    \caption{ (a) Magnetic energy evolution at spatial  resolutions $96^3$, $128^3$ and $256^3$ (Models  AIV, AV and AVI of Table~\ref{tab:models_appendix}). (b) Radial profile of the root-mean-square magnetic field magnitude at $t=48.2~\tau_{A,0}$.
    These simulations are done with $\nu_3=\eta_3=4.4\times10^{-10} R_s^6/\tau_{A,0}$.}
    \label{fig:Emag_reso}
\end{figure}

Finally, we run simulations with the same initial conditions, but at different spatial resolutions: $96^3$, $128^3$, and $256^3$, corresponding to grid spacings $\Delta x =0.046$, $0.035$, and $0.017~R_s$ (models AIV, AV, and AVI of Table~\ref{tab:models_appendix}), respectively. Figure~\ref{fig:Emag_reso}(a) shows the evolution of the total magnetic energy for these three simulations, and Figure~\ref{fig:Emag_reso}(b) shows the radial profile of the root mean squared magnetic field magnitude at $t=48.2~\tau_{A,0}$. There are small differences between these three simulations, but we consider them not to be significant.

\begin{figure}
    \centering
    \includegraphics[width=\columnwidth]{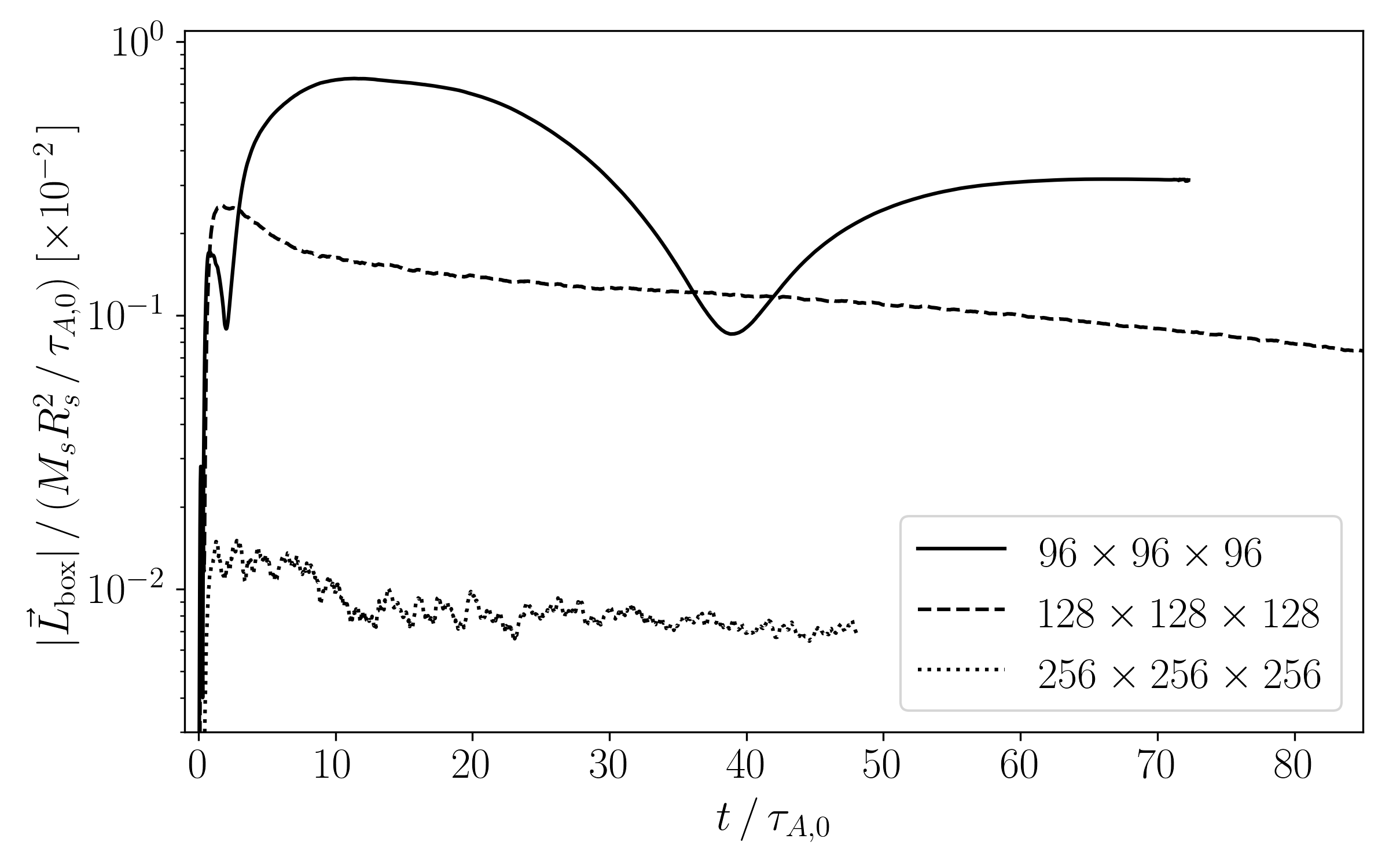}
    
    \caption{Time evolution of the total angular momentum in the box for simulations with $96^3$, $128^3$ and $256^3$ grid elements (Models  AIV, AV and AVI of Table~\ref{tab:models_appendix}).}
    \label{fig:Lbox_reso}
\end{figure}

The {\sc Pencil code} is not a conservative code, i.~e., it conserves quantities only up to the discretization error of the scheme. Thus, we expect that the  conservative properties of the code  would improve with resolution. To verify this, we have plotted in Figure~\ref{fig:Lbox_reso} the evolution of the total angular momentum in the simulation box, calculated as:
\begin{equation}\label{eq:ang_tot}
    \vec{L}(t)=\int_{\rm box} \rho(\vec{r},t) 
    \vec{r}\times\vec{u}(\vec{r},t) \, dV. 
\end{equation}
We neglect the contribution to the angular momentum of the magnetic field since it is much smaller than  the mechanical angular momentum. At late times, the simulation run with resolution $256^3$ has a total angular momentum around two orders of magnitude smaller than the one with a resolution of $96^3$ grid points. The simulations with $96^3$ and $128^3$ grid points developed a rigid rotation inside the star with  angular velocity $\Omega_{\rm rms}=0.38~\tau_{\rm A,0}^{-1}$ and $0.12~\tau_{\rm A,0}^{-1}$, respectively. Although this is a numerical effect, this rotation does not affect the magnetic field evolution, since its period is longer than the Alfv\'en time.

%


\bsp	
\label{lastpage}
\end{document}